\documentclass{aa}
\usepackage{natbib,epsfig,txfonts,graphicx,sidecap}
\bibpunct{(}{)}{;}{a}{}{,}

\def\Ha {H$_\alpha$\,}

\def\Hdelta{H$_\delta$\,}

\def\heiid {He~II~$\lambda$4686\,}
\def\heiic {He~II~$\lambda$4541\,}
\def\heiib {He~II~$\lambda$4200\,}
\def\heiia {He~I+II~$\lambda$4026\,}
\def\heiaa{He~I~$\lambda$4143\,}

\def\heib{He~I~$\lambda$4471\,}

\def\niiie{N~III $\lambda\lambda$4634-4640-4642\,}

\def\ciiie{C~III $\lambda\lambda$4647-4650-4652\,}
\def\Mdot {$\dot M$\,}

\def\kms {km~s$^{\rm -1}$\,} 
\def\Rstar {$R_\star$\,}

\def\Teff {$T_{\rm eff}$\,}

\def\vsini {$v \sin i$\,}
\def\logg {$\log g$\,}

\def\Vr {$V_{\rm r}$\,}

\def\Msun {$M_\odot$\,}
\def\Rsun {$R_\odot$\,}

\def \ao{\frac{1}{\alpha'}}

\begin{document}
\title{Spectroscopic and physical parameters of Galactic O-type 
stars.}\subtitle{I. Effects of rotation and spectral resolving 
power in the spectral classification of dwarfs and giants
\thanks{Based on observations collected at the European Organisation 
for Astronomical Research in the Southern Hemisphere, Chile, 
under programme ID 072.D-0196}}

\author{N. Markova\inst{1}, J. Puls\inst{2}, S. Scuderi\inst{3}, 
S. Sim\'on-D\'iaz\inst{4,5}, A. Herrero\inst{4,5} }
\offprints{N. Markova,\\ \email{nmarkova@astro.bas.bg}}

\institute{Institute of Astronomy with  NAO, BAS, 
	P.O. Box 136, 4700 Smolyan, Bulgaria\\ 
	\email{nmarkova@astro.bas.bg}
\and  Universit\"{a}ts-Sternwarte,
	Scheinerstrasse 1, D-81679 M\"unchen, Germany\\
	\email{uh101aw@usm.uni-muenchen.de}
\and INAF - Osservatorio Astrofisico di Catania, 
	Via S. Sofia 78, I-95123 Catania, Italy\\ 
	 \email{scuderi@oact.inaf.it}
\and Instituto de Astrof\'isica de Canarias, 
      E38200 La Laguna, Tenerife, Spain.\\
\email{ssimon@iac.es}, 
\email{ahd@iac.es}
\and Departamento de Astrof\'isica, 
     Universidad de La Laguna, E-38205 La Laguna, Tenerife, Spain.}
\date{Received; Accepted }
\abstract
{The modern-era spectral classification of O-stars relies on either 
the Walborn or the Conti-Mathys scheme. Since both of these 
approaches have been developed using low-quality photographic 
data, their application to high-quality digital data 
might not be straightforward and be hampered by problems and 
complications that have not yet been appreciated.}
{We investigate the correspondence 
between photographic and digital classification of low luminosity 
O-type stars (dwarfs and giants) of solar metallicity.}
{Using high-resolution spectra obtained with the ESO/MPG 2.2\,m 
telescope in La Silla and following the premises of the Walborn 
and Conti classification schemes, we determined the spectral types
and luminosity classes of 19 Galactic O-type stars and compared 
them to those attributed by Walborn and  Mathys based on 
low-quality data.}
{Our analysis reveals that the $morphological$ spectral types 
assigned using high-resolution data are systematically later 
(by up to 1.5 subtypes) then those attributed by Walborn. By 
means of line-profile simulations, we show that part of this 
discrepancy is more likely caused by the combined effect of 
stellar rotation and high spectral resolution on the depth of 
helium lines used as spectral type indicators. In addition, we 
demonstrate that at least for narrow-lined stars  
the ``rotational effect'' does not disappear when 
the high-resolution spectra are degraded  to the resolution 
of the Walborn standards. We also find evidence of a systematic 
difference between our high-resolution $quantitative$ spectral 
types and those assigned by Mathys.}
{Rotation and spectral resolution are important third parameters 
in the spectral classification of O-type stars. To obtain reliable 
spectral classes within the Walborn approach, the unknown and the 
standard spectra must be compared at the same resolution and \vsini. 
Owing to resolution effects, the Conti approach might also need to be 
updated. } 

\titlerunning{Spectral classification of Galactic O-stars}
\authorrunning{N. Markova et al.}

\maketitle

%

\section{Introduction}

Despite their scarcity, massive stars play an important role 
in the history of the Universe. They are the main engines  
driving the chemical and dynamical evolution of galaxies, enriching 
the interstellar medium with heavy elements, creating H\,{\sc ii} 
regions, and exploding as supernovae. In the distant Universe, 
they dominate the integrated UV radiation in young galaxies. 
Massive stars are possibly key objects for studying and understanding 
exciting phenomena such as the cosmic reionisation and 
$\gamma$-ray bursters. 

In the past decade, our knowledge of the physics of massive stars
has drastically improved because of important developments in 
massive star modelling (both in their interiors and outer envelopes), 
and continuously higher quality observations collected from 
the ground and space. Despite the tremendous progress made so
far, 
a number of new challenging issues (regarding both theory and 
observations) have emerged that need to be addressed urgently. 
Among these are the `weak wind problem' in O-type stars, the wind 
momenta of Galactic B-supergiants being significantly lower than 
predicted, the behaviour of mass loss at the bi-stability jump, the 
effects of rotation and magnetic fields, and wind clumping. 
For a comprehensive review on the status-quo of massive star research, 
with emphasis on radiatively driven mass loss, the interested reader 
is referred to \citep{PVN08}.

For the particular case of O-stars in our Galaxy, the investigation 
of the aforementioned open problems, especially those related to stellar 
winds, requires a large 
sample of objects to be analysed to diminish the error caused by 
the uncertain distances. Since Galactic O-type stars constitute 
an important  reference frame to investigate metallicity effects, 
this point is also crucial for extragalactic surveys. Though 
easy accessible with small and medium size telescopes, Galactic 
O-stars have not been widely observed to derive consisitent stellar 
and wind parameters from quantitative spectroscopy, and no 
more than 50 objects have been studied in detail 
\citep{Herrero02, repo, bouret05, martins05b, dufton06, marcolino}. 
\begin{table*}
\begin{center}
\caption[]{Sample stars along with spectral classification. 
Low-resolution $quantitative$ and $morphological$ classification 
from Mathys \citep{mathys88} and from Walborn \citep{walborn72, 
walborn73, walborn02} and Garrison, Hiltner and Schild (1977, GHS), 
respectively; high-resolution 
$quantitative$ classification from the present study together 
with $morphological$ classification based on the original data 
($R$=48\,000) and degraded ones ($R$=4000). 
Additional 
information about \vsini\, (in \kms; macroturbulence not accounted 
for) and binarity status of the stars is also provided. Binarity 
status from speckle interferometry \citep{mason09} and spectroscopy 
(previous and present).  References for \vsini\,- estimates: H
= \citet{howarth97}; M = Markova et al. 2011 (Paper II); P =
\citet{penny96}; U = \citet{UF}; CE = \citet{CE77}. See text for 
further information.} 
\label{photom}
\tabcolsep1.8mm
\begin{tabular}{lllllrllll}
\hline
\hline
\multicolumn{1}{c}{Star}
&\multicolumn{4}{l}{Previous}
&\multicolumn{1}{c}{\vsini}
&\multicolumn{4}{l}{Present study}
\\
\cline{2-5} \cline{7-9}
\multicolumn{1}{c}{}
&\multicolumn{1}{l}{Walborn}
&\multicolumn{1}{l}{Mathys}
&\multicolumn{1}{l}{GHS}
&\multicolumn{1}{c}{Intfr.$^{a}$/Spst.$^{b}$}
&\multicolumn{1}{c}{}
&\multicolumn{1}{l}{Morph.($R_{48000}$)}
&\multicolumn{1}{l}{Quant.}
&\multicolumn{1}{l}{Morph.($R_{4000}$)}
&\multicolumn{1}{c}{Spect.$^{b}$}
\\
\hline                                
HD~64568       &O3 V((f*))   &	          &&VS        &100, M  &O3 V((f*))  &         &O3 V((f*))  &SS \\
HD~93204       &O5 V((f))    &O5.5 V((f)) &&VS        &115, M  &O6 V((f))   &O5.5 V   &O6 V((f))   &SS \\
HD~93843       &O5 III(f)    &O5.5 III(f) &&VS        &100, P  &O6 III(fc)  &O5.5 III &O5.5 III(fc)&SS \\
CPD~$-$59\,2600&O6 V((f))    &	          &&VS        &142, H  &O6 V((f))   &O6.5 V/III&O6 V((f))   &SS \\ 
HD~63005       &             &&O6 V((f))  &	      & 74, H  &O7.5 V((f)) &O6.5 V   &O7 V((f))   &SS \\
HD~152723      &O6.5 III(f)  &O7 III(f)   &&VD, SB1?  &123, P  &O6.5 III(f) &O6.5 III &O6.5 III(f) &SB1\\
HD~93160       &O6 III(f)    &	          &&VS, SB1?  &205, U  &O7 III(f)   &O7 V     &O7 III(f)   &SB1\\
HD~94963       &O6.5 III(f)  &	          &&VS        &90, CE  &O7.5 II(f)   &O7-7.5 I/III &O7.5 III(f) &SS\\
CPD\,$-$58\,2620 &O6.5 V((f))&	          &&VS        &60, M   &O8 V((f))   &O7 III   &O7.5 V((f)) &SS \\
HD~69464       &O6.5 Ib(f)   &	          &&VS        &82, P   &O7.5 II(f) &O7-7.5 III &O7 III(f)   &SB2?\\
HD~93222       &O7 III(f)    &	          &&VS        &77, P   &O8 III (f)  &O7 III   &O7.5 III(f) &SS\\
HD~91824       &O7 V((f))    &O7 V ((f))  &&VS        &65, H   &O8 V ((f))  &O6.5 V   &O7 V ((f))  &SS\\
CD\,$-$43\,4690&             &O7.5 III(f) && 	      &120, M  &O7 III(f)   &O6.5 V/III &O7 III(f)   &SS \\
HD~92504       &O8.5 V((n))  &O9 V        &&  	      &200, U  &O8.5 V      &O9 III/V &O8.5 V      &SS \\
HD~151003      &O9 II	     &O9.5 III    &&VS, SB1?  &120, U  &O9 II       &O9 III   &O9 II       &SB1 \\.   
HD~152247      &O9.5 II/III  &O9.5 III    &&VS, SB2   &120, P  &O9 II       &O9 III   &O9 II       &SB2 \\
HD~302505      &             &&O8.5~III   &           & 80, M  &O9 III      &O8 III   &O9 III      &SS \\
CPD\,$-$44\,4865& 	     &O9.7 III    & &	      & 80, M  &B0 III      &O9.5 III &B0 III      &SS \\
HD~69106       &             & &B0.5 IVnn &           &325, H  &B0.2 V      &O9.7     &B0.2 V      &SS \\
\hline
\end{tabular}
\end{center}
$^{a}$ Results from the speckle interferometric survey 
of \citet{mason09}: VD = visually double; VS = visually single object.\\
$^{b}$ Spectroscopic binarity status: SS = spectroscopically single 
object; SB1/SB2 = single-/double-lined spectroscopic 
binary; SB1/2? = possible spectroscopic binary.\\
 
\end{table*}

The main goal of our project is to increase the number of Galactic 
O-type stars with reliably determined physical parameters, using 
high-quality spectra and applying the methods of quantitative 
spectroscopic analyses. On the basis of new data and incorporating 
similar data from previous investigations, we plan to reinvestigate 
the latest calibrations for these stars \citep{martins05a}, to address 
the important question of macroturbulence and its effect on the 
derived \vsini, and to investigate the weak wind problem 
in stars of solar metallicity. Since the {\it absolute} calibration 
of the various relationships between physical parameters and observed 
spectral characteristics (e.g., line strengths or, equivalently, 
spectral types) requires reliable estimates of all involved quantities, 
a thorough investigation of the accuracy of the spectral types and luminosity 
classes initially assigned to our targets and an update of their 
binary/multiplicity status are required.

In this first paper of the series, the main results from the spectral
classification of the low luminosity stars in our sample (dwarfs and 
giants) are presented. A second paper providing the physical properties 
of these stars derived by means of the latest version (V10.1) of 
the state-of-the-art model atmosphere code {\sc fastwind} \citep{puls05} 
is currently in preparation and will be published soon (Markova et al. 
2011, hereafter Paper II). A third paper concentrating on the spectral 
classification and model atmosphere analysis of the supergiants in 
our sample is planned for the near future.

The present paper is structured as follows. In Section~\ref{obs},
 we describe the stellar sample and the observational material 
 underlying the project. In Section~\ref{spec_class}, we outline
the main steps in our classification procedure. Section~ \ref{binary} 
deals with the possible binarity/multiplicity among
 the sample stars. In 
Section~\ref{results}, the correspondence between our 
high-resolution $morphological$ and $quantitative$ classification and 
the classifications attributed by Walborn \citep{walborn72, 
walborn73} and \citet{mathys88} for stars in common is investigated, 
and a possible explanation of the derived discrepancies is provided. 
The spectral types determined in the framework of the Walborn and 
Conti-Mathys schemes from our high-resolution observations are compared 
in Section~\ref{quant-morph}. Finally, in Section~\ref{conclusions}, 
we summarise the main results of our study and comment on some 
implications these findings might have in the future. Extensive 
comments on each sample star and an atlas of the corresponding 
high resolution spectra 
are presented in the appendix.

\section{Observations and data reduction}
\label{obs}

The total sample consists of 40 bright ($V$~$\le$~10~mag) O-type 
stars in the Milky Way originally classified as dwarfs, giants, 
and supergiants, with spectral types ranging from O3 to B0.5. 
The targets 
were observed with the FEROS spectrograph \citep{kaufer99} at the 
ESO/MPG 2.2\,m telescope in La Silla on the 4th and 5th of February, 
2004. Each spectrum covers a wavelength range from about 350 to 
about 920\,nm with a spectral resolution $R$ = 48\,000. Exposure 
times ranging from 400 to 2500\,s were used to obtain a typical 
signal-to-noise ratio (hereafter S/N) of 150-200 per resolution 
element. One-dimensional, wavelength-calibrated spectra were 
extracted using the FEROS pipeline. 

For our present study, we considered only the subset of the 
sample dwarfs and giants (19 in total) to diminish the 
effects of stronger winds on the outcome of the classification 
analysis. The stellar IDs along with our present and previous  
spectral classifications by Walborn \citep{walborn72, walborn73, 
walborn02} and \citet{mathys88} are presented in Table~\ref{photom}. 
We note that for three sample stars that have  been classified by 
neither Walborn nor Mathys, previous classifications based on  
low-resolution photographic spectra and MK standards were adopted  
(Column 4 of Table~\ref{photom}).

\section{General comments}
\subsection{Spectral classification}
\label{spec_class}

The modern-era O-star spectral classification relies on either 
the Walborn or the Conti-Mathys schemes, both being somewhat 
related to the MK system defined essentially in terms of standard 
stars. In particular, based on a careful investigation of MK 
standards observed at a spectral resolution twice that of the 
MK atlas, Walborn (1971, 1972, 1973) developed a two-dimensional 
empirical system for O-type stars of solar metallicity. This system 
was additionally worked out by Walborn and coworkers, and transferred 
to digital data of similar \citep{wf90} or higher \citep{walborn02} 
resolution. Since the Walborn approach relies on a visual inspection 
of the spectra, and the visibility of certain lines and certain
line ratios are used as criteria, it is usually referred to as a
$morphological$ classification. Independently of and simultaneously 
with the morphological approach, Conti and collaborators \citep{conti71, 
CL74, CF77} developed an alternative method based on logarithmic ratios 
of equivalent widths (EW) of certain helium and metal lines, calibrated 
against MK spectral types. The $quantitative$ classification system 
of Conti was refined by Mathys \citep{mathys88, mathys89}. 

To classify our targets, we applied both the Walborn and the 
Conti-Mathys systems. In the former case, we considered the visual 
impression of our high-resolution spectra and, following the 
premises of \citet{wf90}, performed a $morphological$ classification.
However, since the line visibility depends directly on the 
data quality, we refrained from using criteria related to 
the initial appearance of certain lines and concentrated only on 
{\it eye-estimated} line ratios, to avoid possible systematic 
effects caused by the higher resolution (and S/N) of our spectra. 
We primarily used the He~I to He~II line ratios (particularly 
those of both \heiia\, to \heiib\, and \heib\, to \heiic) to determine 
the spectral subtype, except for the hottest stars where criteria
based on the nitrogen ionization equilibrium \citep{walborn02} were 
instead applied. For the luminosity class, we followed 
the Walborn approach and used different criteria at different 
subtypes. For example, at earlier types we considered the effect 
of luminosity on the strength of both the N~IV~$\lambda$4058 and \heiid\, 
emission, and the N~V~$\lambda\lambda$4604, 4620 absorption. 
At spectral types from O6 to O8, the main luminosity indicators 
are \heiid\, and \niiie: in dwarfs, the former line appears strongly 
in absorption, accompanied by weak N~III emission - the V((f)) category; 
in giants, the \heiid\, absorption weakens and can even vanish while 
the \niiie\, emission strengthens - the III(f) category. At types 
O9-B0, the primary luminosity criterion is the strength of the Si~IV doublet
around \Hdelta, whose absorption increases with luminosity.

Our $quantitative$ classification has been performed by exploiting 
the logarithmic ratios 
$\log W'$ = $\log EW$(HeI$\lambda$4471) $-$ $\log EW$(HeII$\lambda$4542), 
which un\-am\-biguously determines the spectral type, and
$\log W''$ = $\log EW$(SiIV$\lambda$4089) $-$ $\log EW$(HeI$\lambda$4143), 
which is the main luminosity indicator for stars of spectral types 
O7 and later \citep{conti71}. For stars earlier than O7, on the 
other hand, the negative luminosity effect in  \heiid  with a 
demarcation line between class V and III set at 
$\log~EW$(HeII4686) =$-$0.25\footnote{ $\log~EW$(HeII4686) larger 
than -0.25 determines the luminosity class V.} was used
(\citealt{mathys88} and references therein). 
The equivalent widths of the lines involved in the $quantitative$ 
classification  are provided in the appendix (Table~\ref{EW}). 
The accuracy of these estimates ranges from 0.02 to 0.07~dex, 
with lower values being typical of stronger lines. The  
errors in $\log~W'$ and  $\log~W''$ are  between 0.05 and 0.07~dex 
(i.e., less than half a subtype), and about 0.1~dex, respectively.  

Our final $morphological$ and $quantitative$ classifications 
are listed in Cols. 7 and 8 of Table~\ref{photom}. 
Column 6 gives \vsini-estimates as adopted 
in the present study, primarily drawn from \citet{penny96} and 
\citet{howarth97}\footnote{With respect 
to our targets, the consistency between the two datasets is 
good,  to within $\pm$10~\kms.}, if present. Otherwise, individual data  
provided by \citet{UF} or derived in Paper II were used instead. 

\subsection{Spectroscopic binarity} 
\label{binary}

It is well known that spectroscopic binarity can significantly 
bias the classification of stars, modifying the visual appearance of 
the spectrum and altering the observed equivalent widths of lines 
used as spectral type and luminosity class diagnostics. This point is 
particularly important in the case of hot massive stars where the 
number of double/multiple systems seems to be very large (e.g., 
\citealt{mason09} and references therein). 

To investigate the binary status of our targets, we checked 
the literature for known/suspected binaries among our sample 
stars and found that three of them -- HD~93160, HD~151003 
\citep{gies87}, and HD~152723 \citep{full96} -- have been 
$suspected$ to be single-lined spectroscopic binaries (SB1), 
because of their \Vr-variability. Large variations in \Vr\ 
were also detected for another star, HD~64568 \citep{SN86}. 
Since \Vr - variability can be caused by binary motion, but 
may also originate from stellar pulsations and/or wind effects,
 we decided to keep the \Vr-variables in our target list for 
further clarification. Meanwhile, we rechecked the literature 
for newer results and found that one of our sample stars 
(HD~152247) had been recognized as a double-lined spectroscopic 
 binary \citep{sana08}; another one (HD~152723) was resolved 
as a close double system via $V$-band speckle interferometry, 
while for other 12 objects (flagged with ``VS'' in column 5 
of Table~\ref{photom} ) a ``null companion detection'' in the 
angular separation range 0."035~$\le\rho\le$1."5 and $\Delta m \le$ 3~mag 
was reported (\citealt{mason09} and references therein). We kept the 
two binaries in the present sample with the primary goal of 
investigating the effect of binarity on the outcome of our spectral 
classification. 

We also tried to constrain the spectroscopic binarity status 
of all sample stars by our own. To this end, we adopted the following 
criteria as indications for possible binarity: (i) composite spectrum; 
(ii) discrepant shifts, widths and/or strengths of lines in the spectrum; 
(iii) large differences in \Vr\, and \vsini\, when measured at different 
observational epochs; (iv) periodic variations in photometry and/or radial
velocity; and (v) discrepant spectral classification from various observational 
epochs. Since physical phenomena that differ from binarity 
can give rise to periodic photometric and \Vr-variability while discrepant 
spectral classifications might be due to third parameter effects (see below), 
these last two criteria were considered only as suggestive. 

A comprehensive description of our findings for each star is presented 
in the appendix. The spectroscopic status of the targets is listed in 
Column 10 of Table~\ref{photom}: ``SS'' denote spectroscopically single 
objects, ``SB2'' are double-lined, and ``SB1'' single-lined spectroscopic 
binaries. 

In summary, following our approach we confirm the binary status 
of HD~152723 and HD~152247, provide strong evidence of a companion in 
HD~93160 and HD~151003, and suspect duplicity in HD~69464. 

\section{Comparison with previous classifications from 
low resolution spectroscopy}
\label{results}
\subsection{The present high-resolution morphological 
classification and the Walborn results}
\label{firstcomp}

To investigate whether the results derived by means of the 
Walborn scheme might be influenced by the spectral resolution, 
we 
\begin{figure}
\center
\resizebox{\hsize}{!}
{\includegraphics[width=6cm,height=4cm]{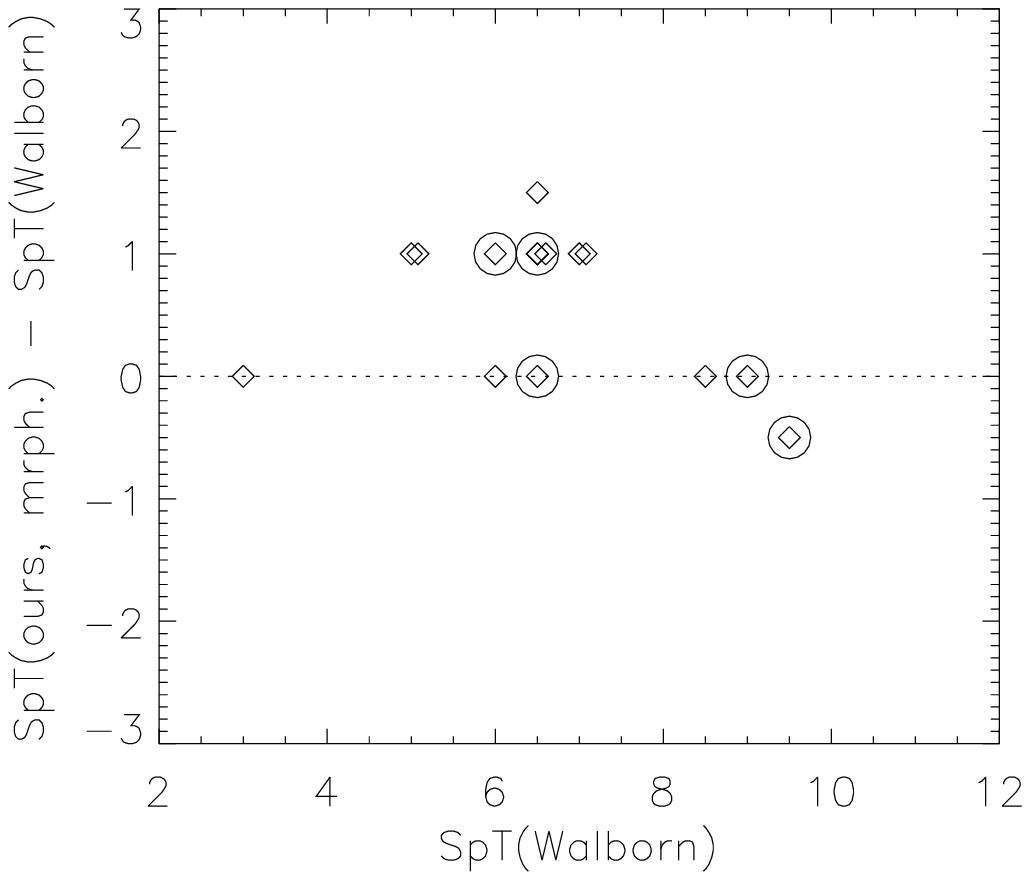}}
\\
\vspace{-0.5cm}
\resizebox{\hsize}{!}
{\includegraphics[width=6cm,height=4cm]{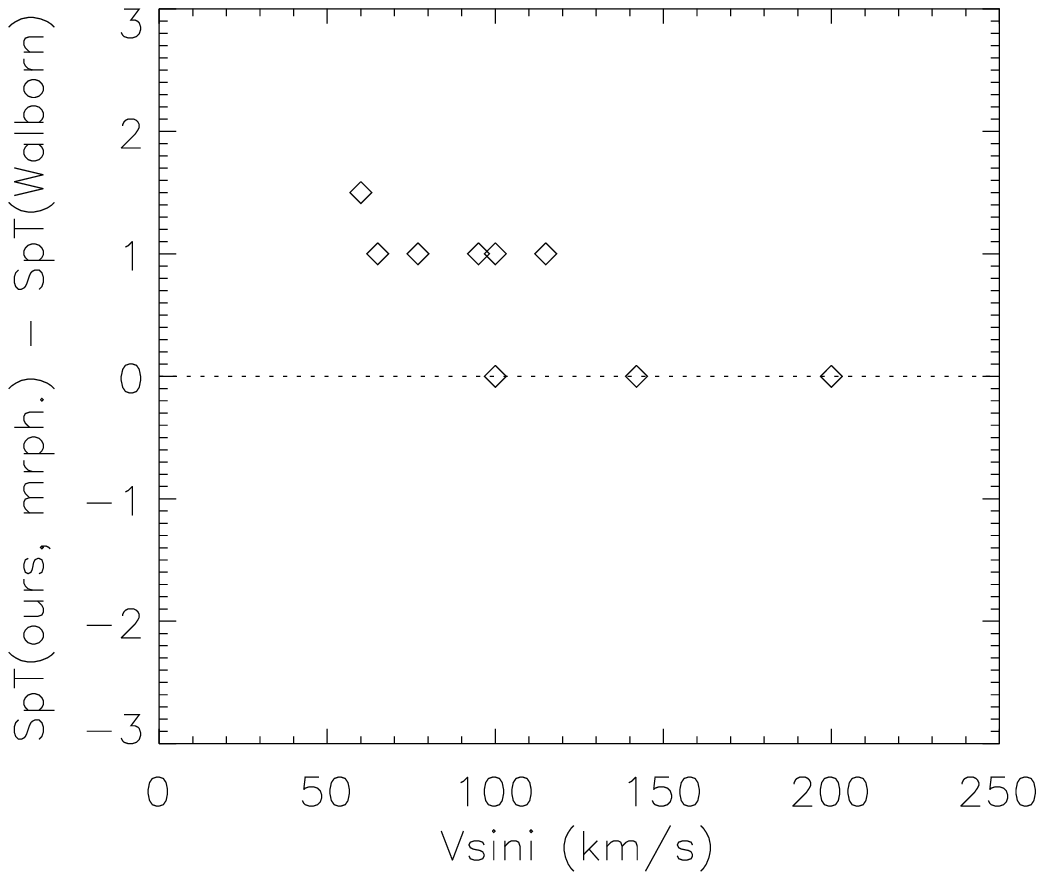}}
\caption{{\it Top}: Comparison of our high-resolution {\it
morphological} spectral types and those attributed by Walborn by 
means of low-resolution $photographic$ spectra ($\Delta\lambda$ = 1.2 \AA), 
for 14 stars (dwarfs and giants) in common. The X-scale corresponds 
to spectral types from O2 to B2. Stars considered/suspected to be 
SBs are additionally marked with large circles. {\it Bottom}: The 
spectral type discrepancy as a function of \vsini. To eliminate 
binarity effects on the observed \vsini, stars considered/suspected 
to be SBs have been discarded.  Note that the adopted \vsini-estimates 
do not account for the effects of macroturbulence.
}
\label{SPT_morph}
\end{figure}
compared our $morphological$ spectral types with those attributed 
by Walborn, based on low-resolution ($\Delta\lambda$ = 1.2 \AA) 
$photographic$ spectra, for 14 stars in common (see Column 2 of 
Table~\ref{photom}). As illustrated in Fig.~\ref{SPT_morph} 
(top panel), the high-resolution $digital$ classification 
tends to result in somewhat later (by up to 1.5 subtypes) 
spectral types than those derived by Walborn. This discrepancy 
is larger than the error in the $morphological$ classification 
(typically less than one subtype), thus significant.
An interpretation in terms of variable wind effects or possible 
binarity does not seem likely, since all targets are low luminosity 
objects, i.e., have low-density winds, and the majority of them 
do not appear to have a companion. In addition, the 
distribution of the data suggests that the derived discrepancy might 
be present at intermediate subtypes only. Since the main spectral type 
indicator at these subtypes is the {\it eye-estimated} ratio of either 
\heiia\, to \heiib, or \heib\, to \heiic, this result implies in 
turn 
that within the high-resolution digital spectra either \heiia\, and 
\heib\, appear to be $visually$ stronger or \heiib\, and
\heiic\, appear to be $visually$ weaker (or both), compared to their
corresponding strengths as estimated from the $photographic$ spectra.

Apart from stars with discrepant spectral types, our analysis 
shows that there are also others with identical $morphological$ 
classifications as derived from high- (digital) and low-resolution 
(photographic) spectra. One of these (HD~64568) is a very early 
O3 dwarf (see below); two other stars (HD~152723, O6.5 and 
HD~151003, O9) are considered to be spectroscopic binaries 
(present study); and the final two are among the fastest rotators 
in our sample (CPD -59~2600, O6, \vsini=142~\kms and HD~92504, O8.5, 
\vsini=200~\kms). The latter finding is quite intriguing since it 
might imply that the encountered discrepancy is also related to
rotation. We test this possibility in Fig.~\ref{SPT_morph} (lower panel),
where the differences in spectral type are plotted against \vsini.
At least for our subsample, the spectral type discrepancy indeed 
refers to stars with relatively slow rotation (\vsini$\le$120~\kms). 
\footnote{Similar results apply for the three stars classified by
\citet{GHS77}, see Table~\ref{photom}, Column 4.} The only exception 
is HD~64568 with \vsini = 100~\kms, which, however, has not been
classified in terms of the {\it eye-estimated} ratios of He~I to He~II but
instead exploits the relative strengths of nitrogen lines of different
ionisation. 

Summarizing, 
our analysis shows that using high resolution
$digital$ spectra of O-type stars of solar metallicity results in 
somewhat later $morphological$ subtypes, where this discrepancy is 
more pronounced in narrow-lined stars.

\begin{figure}
\center
\resizebox{\hsize}{!}
{\includegraphics[width=8.5cm,height=5cm]{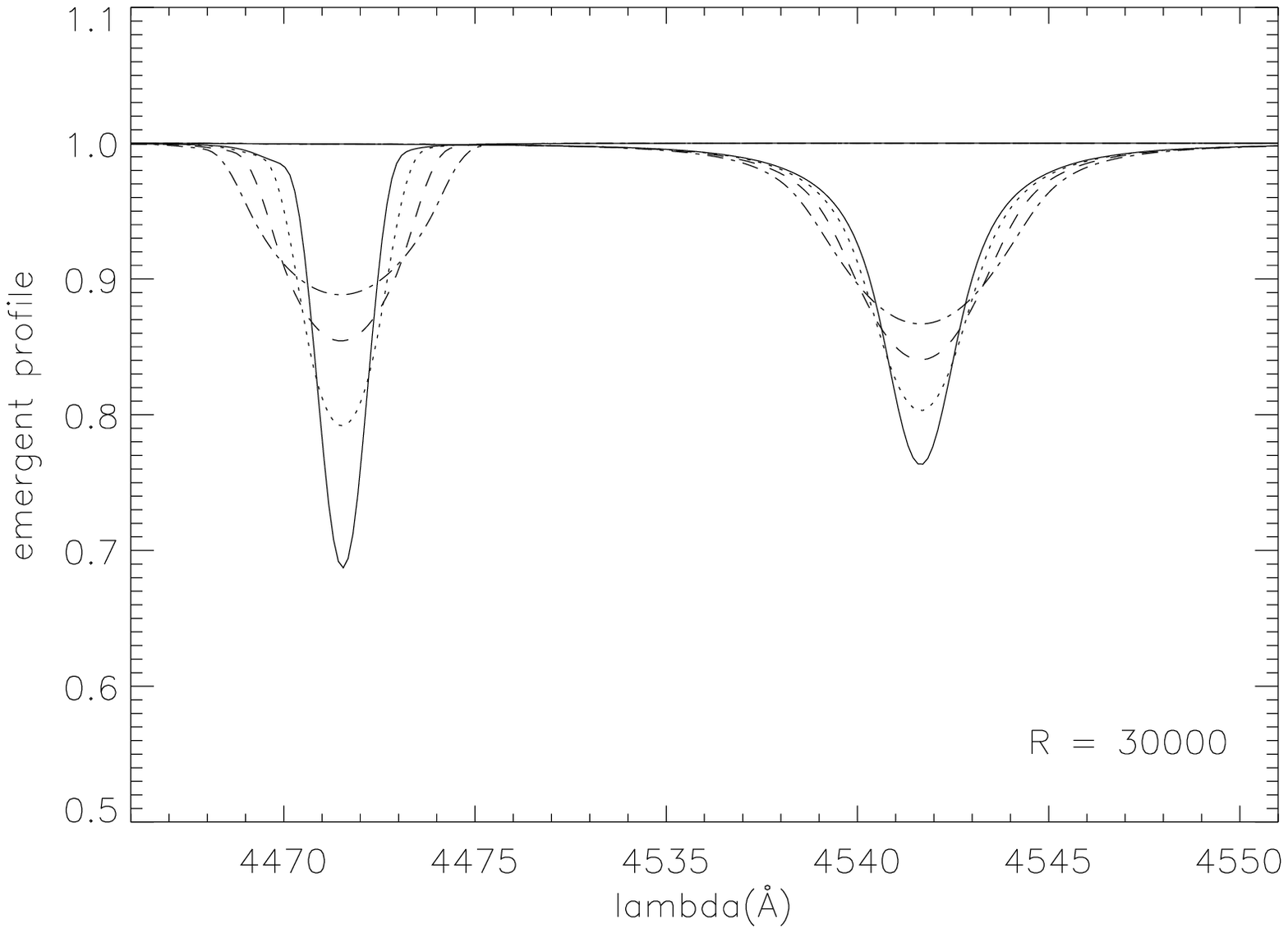}}\\
\resizebox{\hsize}{!}
{\includegraphics[width=8.5cm,height=5cm]{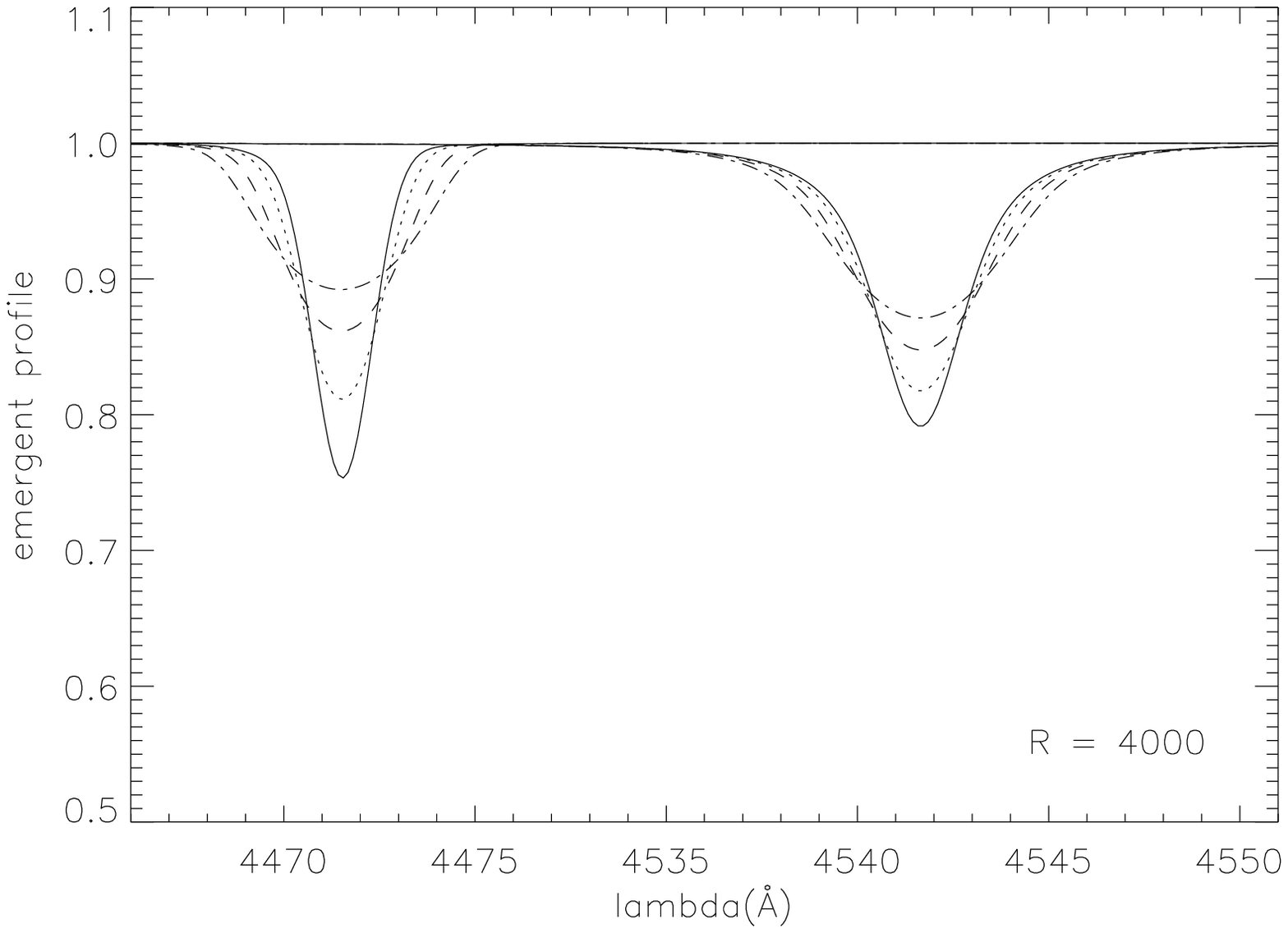}}
\caption{Examples illustrating the sensitivity of the central 
depth of \heib\, and \heiic\, to stellar rotation (\vsini=60 
(solid), 100 (dotted), 150 (dashed), 200 (dash-dotted) \kms), and 
spectral resolution ($R$ = 30\,000 (top), 
4\,000 (bottom)). Model profiles from {\sc fastwind}, calculated 
at \Teff=38~kK, \logg=3.7 and log~$Q$=-7.72, with
log~$W'$=-0.19 corresponding to subtype O6.5. 
}
\label{res_rot}
\end{figure}
\begin{figure*}
\center
{\includegraphics[width=18cm,height=7cm]{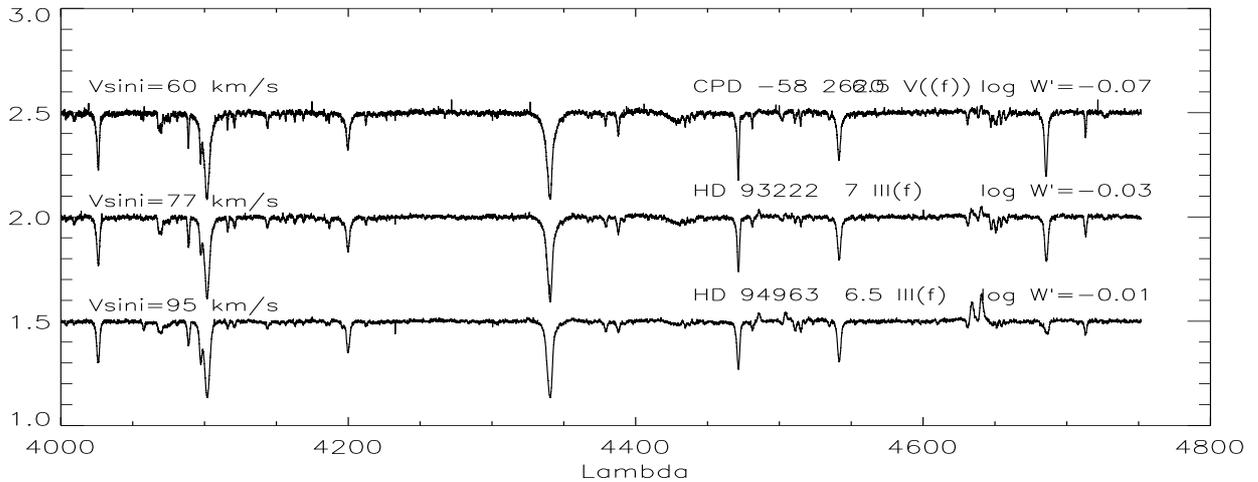}}
\caption{A set of FEROS spectra illustrating the $combined$ effect 
of rotation and high spectral resolution on the central depth of 
the strategic He~I $\lambda$4026, 4471 and He~II $\lambda$4200, 4541 
lines, for stars of similar spectral type (as attributed by 
Walborn) and similar $log W'$, as measured from our high-resolution 
data. Note that in the Walborn scheme the O7 subtype is defined by 
\heib~$\approx$~\heiic, whereas in our spectra \heib~ is always deeper 
than \heiic\ (although its EW is actually smaller, log~$W'\le$ 0), 
indicating a spectral type of O8-8.5.}
\label{spectra}
\end{figure*}

\paragraph{High-resolution against low-resolution} $morphological$ 
luminosity classes. A comparison of the luminosity classes 
determined in this work with those attributed by Walborn for 
14 stars in 
common shows perfect agreement for all but two stars - 
HD~94963 and HD~69464. This 
finding suggests that at least for dwarfs and giants the luminosity 
class criteria within the Walborn scheme are not affected by resolution, 
rotation, and/or origin of data.

\subsubsection{Influence of rotation and spectral resolution}
\label{rot_effect}

{\it That} rotation should influence the depths of He~I and He~II 
lines differently is to be expected from the different intrinsic 
widths of these lines, caused by the Stark effect (linear in 
He~II and quadratic in He~I). This point is illustrated in 
Fig.~\ref{res_rot}, where synthetic profiles for \heib\, 
and \heiic, as calculated from a {\sc fastwind} model at \Teff=38~kK, 
\logg=3.7, and log~$Q$=$-$7.72\footnote{$Q$=\Mdot/\Rstar$^{1.5}$ 
is the optical depth invariant for recombination-based diagnostics 
such as \Ha, introduced by \citet{puls96} to characterise the 
strength of the wind. 
log~$Q$=$-$7.72 (\Mdot in \Msun/yr, \Rstar in \Rsun) represents 
the case of a low density wind, originating in a low luminosity 
O-type star.}, have been convolved with four values of \vsini, 
while keeping the spectral resolution fixed at $R$=30\,000 (top panel) 
and 4\,000 (bottom panel). The main result of our calculations  
is that  at the same temperature, surface gravity, chemical 
abundances, mass loss rate, and spectral resolving power the 
{\it eye-estimated} ratio of \heib to \heiib can indicate different spectral 
types, depending on \vsini. A comparison of the profiles shown in 
Fig.~\ref{res_rot} indicates that the $morphological$ spectral 
types  for fast (\vsini$\ge$150~\kms) and slowly (\vsini$\le$100~\kms) 
rotating stars can differ by about half (at $R$=4\,000) to about 
one and a half (at $R$=30\,000) subtypes, where slowly rotating 
stars display a later subtype.
Similar results have been obtained for \heiia\, and \heiib\, 
determining the $morphological$ spectral types from O5 to O6.5.

Illustrative examples of the $combined$ effect of rotation and high 
spectral resolution on the visual strength of strategic He~I and 
He~II lines are shown in Fig.~\ref{spectra}, where the FEROS spectra 
of three narrow-lined sample stars, originally classified as 
O6.5 - O7 by Walborn, are displayed along with log~$W'$-values from 
our high-resolution EW measurements.  The central depths 
of \heib\, and \heiic\, lines obviously react in  the way predicted by 
our model calculations, and indicate a somewhat later spectral 
type, namely  O8 to 8.5 depending on \vsini.

Thus, we conclude that the established discrepancy between 
high- and low-resolution $morphological$ spectral types 
(mostly for stars of intermediate subtypes) is likely related 
(at least in part, see below) to the interplay between 
stellar rotation and spectral resolution leading to different 
depths of He~I and He~II lines, where the profiles become 
deeper at lower rotational speeds and higher resolution. We
emphasize that for stars with identical intrinsic 
parameters but different rotational speeds, there will be a 
decisive difference in spectral types derived by either 
morphological or quantitative classification: whereas the
morphological types will vary according to rotational (and 
resolution) effects, types derived by a quantitative 
classification will be rather unique, since 
rotational broadening preserves the EW and thus $W'$.

\subsubsection{The modern-era $morphological$ classification.}  
\label{spectra_degrading}

It is well known that the overall appearance of a stellar 
spectrum is strongly dependent on its resolving power. Because 
of this, the modern-era morphological classification is 
usually performed using (original) spectra that have been 
degraded to the resolution of the standards. In this 
process, one expects the morphology of the degraded spectra to 
closely resemble that of the low resolution classification standards. 
To our knowledge, however, this expectation has not been 
so far checked using real data.

\begin{figure}
\resizebox{\hsize}{!}
{\includegraphics[width=6cm,height=4cm]{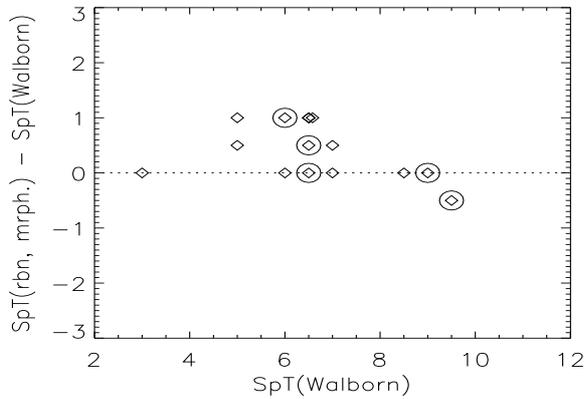}}
\caption{Comparison of our $morphological$ spectral types, 
based on the high-resolution spectra degraded to $R$=4\,000, 
and those attributed by Walborn using low-resolution photographic 
spectra ($\Delta\lambda$~=~1.2 A) for 14 stars in common. The X-scale 
corresponds to spectral types from O2 to B2. Stars considered/suspected 
to be SBs are marked with large circles.} 
\label{rbn}
\end{figure}

For a first test of this issue, the original, high-resolution 
spectra of our targets were degraded to $R$=4\,000
\footnote{which matches the quality of the photographic spectra 
used by Walborn, namely $\Delta\lambda$ = 1.2 \AA.} (using the 
IDL procedure ``rebin.pro''), and subsequently classified following 
the Walborn criteria. In Fig.~\ref{rbn}, we confront the spectral 
types obtained in this way to those attributed by Walborn for stars in 
common. Although the agreement between the two data sets has 
improved (compared to the one shown in the top panel of 
Fig.~\ref{SPT_morph}), the spectral types based on the $degraded$ 
spectra (Column 9 of Table~\ref{photom}) still tend to be somewhat 
later than those derived by means 
of the $photographic$ classification, where again this discrepancy 
is more significant in narrow-lined stars. To check whether this 
finding depends on the procedure used to degrade the spectra, we 
repeated the process using an own IDL procedure based on convolution 
by Fourier-transforms, and obtained similar results. 
\begin{figure*}
{\includegraphics[width=18cm,height=5cm]{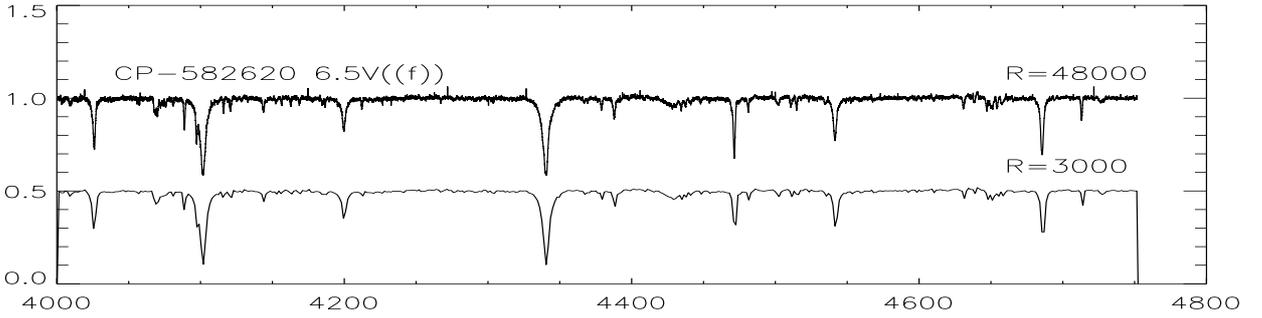}}
\caption{The spectrum of the narrow-lined star CPD-58\,2620 
(\vsini=60~\kms) at the original resolving power $R$=48\,000 
and degraded to $R$=3\,000. Spectral classification from 
Walborn based on low-quality photographic spectra. Note that 
\heib\ appears of similar strength as \heiic in the degraded 
spectrum, indicating the O7 subtype.}
\label{resol_effect_2}
\end{figure*}

The effect of significantly different spectral resolving 
power on the overall appearance of the spectrum is
illustrated in Fig.~\ref{resol_effect_2}, where 
the original, high-resolution spectrum of one sample star 
with relatively narrow lines (CPD -58\,2620, \vsini=60~\kms),
classified as O6.5 by \citet{walborn73}, is shown together with 
its rebinned spectrum\footnote{Here the original spectrum 
has been degraded to $R$=3\,000 ($\Delta\lambda$ = 1.5 \AA) 
to match the quality of the standards shown in the Walborn and 
Fitzpatrick atlas.}. Even at this low resolution,  
the spectrum of CPD -58\,2620 obviously does not resemble that of HD~93146, 
the classification standard of O6.5V, but is similar to that 
of 
15~Mon, the classification standard for O7. Thus, and at least 
for the case of narrow-lined stars of intermediate spectral type, 
it appears that degrading the high-resolution spectra to 
the resolution of the Walborn standards does not necessarily 
guarantee that they will resemble the morphology of the corresponding 
standards, since in this case the comparison objects (the one under 
consideration and the standard) might have different \vsini. The 
larger the difference, the larger the offset in derived spectral
type. 

The results outlined above are somewhat surprising since 
(i) we degraded our spectra to the resolution of 
the photographic data, and (ii) we classified the same stars 
as Walborn, i.e., the 
resolution and rotational effects should be the same in both 
data sets.\footnote{Note that the corresponding spectral types 
might still be biased against the standards due to rotational 
effects, but both (ours and Walborn's) in the same way, see below.} 
Since there are still differences, additional effects must be 
at work. What might these effects be?
\begin{itemize}
\item[i)] The remaining discrepancies shown 
in Fig.~\ref{rbn} might be caused by the intrinsic differences 
in the classification techniques applied in the 
$photographic$ and $digital$ classifications: while in 
the former case the eye responds to the total flux 
transmitted/blocked by the investigated lines (recalling, that 
in the early seventies of the last century the classification 
was performed by viewing the spectrographs through a microscope), 
in the latter it is guided by the central depths of the lines. Consequently, 
both estimates might deviate from each other to some degree, 
particularly at intermediate subtypes where the differences in 
line strength are smaller than in other regions.  
\item[ii)] Since O-stars are often embedded in emitting gas, 
underestimated nebula contamination of He~I lines in the photographic 
spectra might lead to systematically different classifications  
from  those originating from the high-resolution digital 
data.
\item[iii)] Finally, there is a certain chance of line profile 
variability.  Owing to the systematic character of the discrepancy, 
we consider this possibility, however, less likely.
\end{itemize}

Thus, we have to conclude that part of the discrepancies 
between
$morphological$ classifications  attributed using high-resolution 
digital and low-resolution photographic spectra (Fig.~\ref{SPT_morph}) 
is due to some additional effects to resolution and 
rotation alone. This point was also discussed by \citet{walborn10b}. 

\subsection{The present high-resolution quantitative 
classification  and the Mathys results}
\label{mathys}

Since the Conti spectral type classification relies on the 
logarithmic ratio of the measured equivalent widths of \heib\, 
and \heiic, it is essential to investigate the correspondence 
between the present equivalent width scale and that from 
similar studies using low-resolution spectra (e.g., \citealt{conti71}; 
\citealt{CF77}; \citealt{mathys88, mathys89}), to 
provide a useful comparison. We limit ourselves to the 
study by \citet{mathys88}, which has the largest overlap 
with our stellar sample.

\begin{figure}
\resizebox{\hsize}{!}
{\includegraphics[width=6cm,height=4cm]{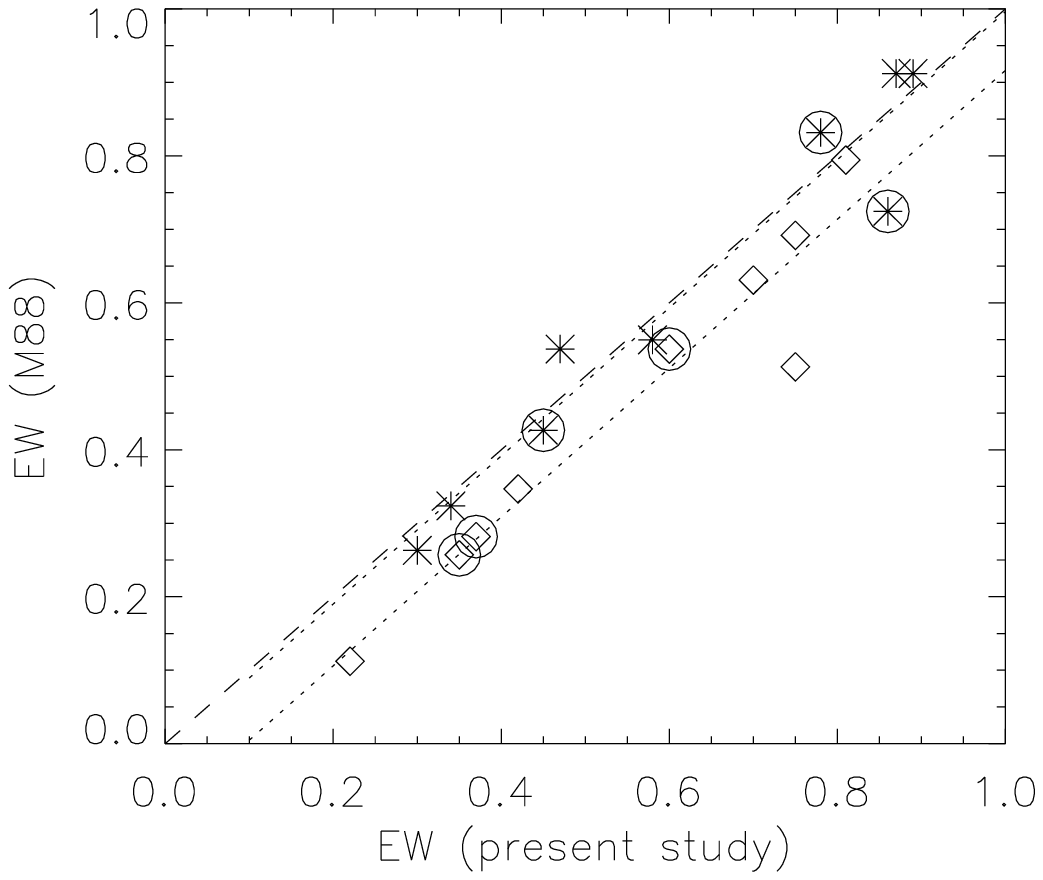}}
\resizebox{\hsize}{!}
{\includegraphics[width=6cm,height=4cm]{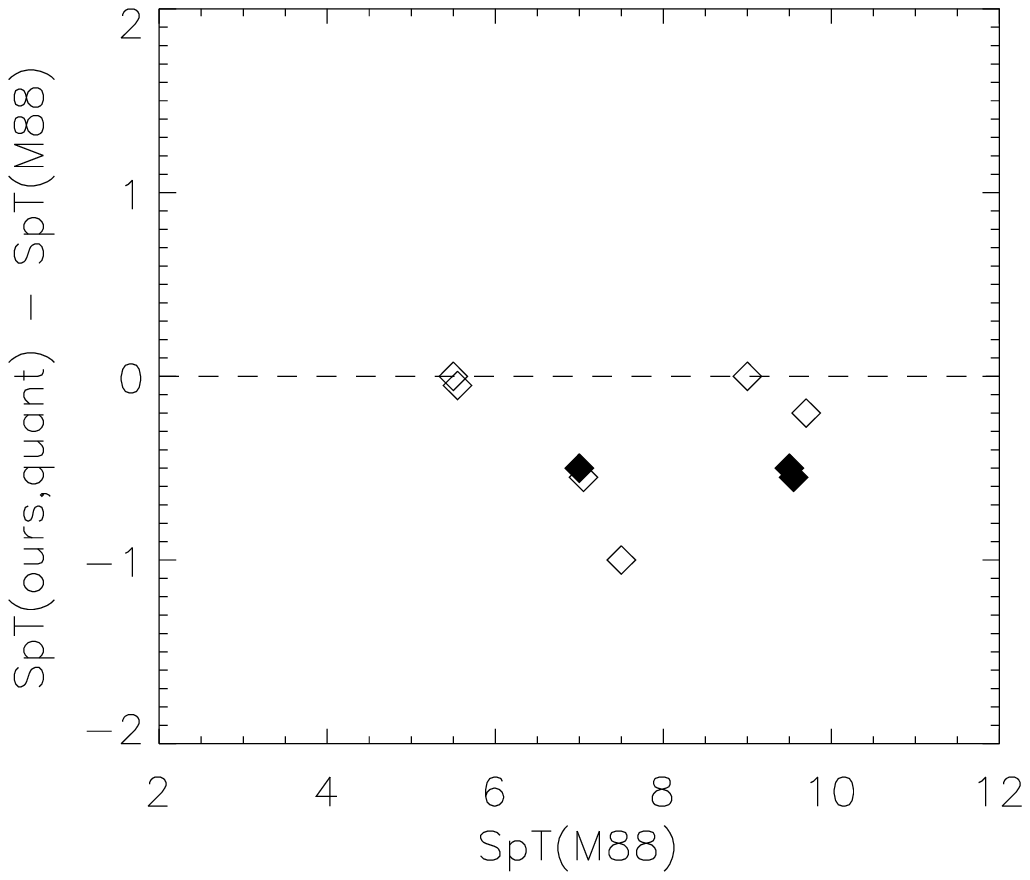}}
\caption{$Top$: Comparison of our high-resolution 
equivalent widths (in \AA) of \heib (asterisks) and He~II~$\lambda$4541
(diamonds) with those from \citet{mathys88}, for nine stars in common. 
Dotted lines represent linear fits to the corresponding data. Note 
the non-zero offset in the EW scale of \heiic.
$Bottom$: Comparison of the corresponding high-resolution 
$quantitative$ spectral types. Spectroscopic binaries are highlighted 
by either  large circles (top panel) or filled diamonds (bottom panel).}
\label{comp_EW}
\end{figure}

In the top panel of Fig.~\ref{comp_EW}, our EW measurements 
for He~I $\lambda$4471 (asterisks) and \heiic (diamonds) are 
compared to those from \citet{mathys88} for nine stars in common. 
Despite the small number of objects, a clear trend for \heiic 
is visible, indicating that our measurements result in 
systematically stronger EWs than derived by Mathys. A linear fit 
to the corresponding data confirms this notion, displaying a 
non-zero offset in the EW scale (in \AA) for this line given by\\
\centerline{
\footnotesize{EW4541(ours) = ($-$0.096$\pm$0.06) + (1.01$\pm$0.11) $\cdot$
EW4541(M88).}}
Interestingly, no indication of any shift was found in the EW scale of
He I $\lambda$4471\\
\centerline{
\footnotesize{EW4471(ours) = ($-$0.01$\pm$0.06) + (1.01$\pm$0.1) $\cdot$ 
EW4471(M88).}}
These results imply that at least some of our targets might appear 
as a somewhat earlier spectral type than denoted by \citet{mathys88}. 
The data  shown in Fig.~\ref{comp_EW} (lower panel) indicate
that our spectral types tend to be earlier, by half a subtype. 
Though this discrepancy is comparable to the error budget of our 
$quantitative$ classification, it might be significant because 
of its 
systematic character (see Section~\ref{conclusions}).

Concerning luminosity classes, perfect agreement between the present 
$quantitative$ luminosity classes and those attributed by \citet{mathys88} 
was established for the nine stars in common. 
 
The discrepancy between our present and the Mathys EW scale for 
\heiic is difficult to interpret. On the one hand, binarity cannot 
be an issue, because both double and single stars are involved. 
On the other hand, a higher precision of our measurements can be 
expected, because of the higher quality of our spectra. This higher 
quality will allow us to consider the contribution of the broad Stark 
wings more correctly. Given that in \heiic\, the wings are more 
extended than in \heib\, (linear against quadratic Stark effect), 
we speculate that the effect of a higher $R$ and S/N on the
measured EWs will be more pronounced in the former than in the 
latter line.

To conclude, the luminosity class criterion in the Conti 
scheme is rather insensitive to changes in $R$, while the spectral 
type indicator seems to be sensitive, though at a (very) low level.

\section{Correspondence between quantitative and morphological 
spectral types} \label{quant-morph}

\begin{figure}
\resizebox{\hsize}{!}
{\includegraphics[width=6cm,height=4cm]{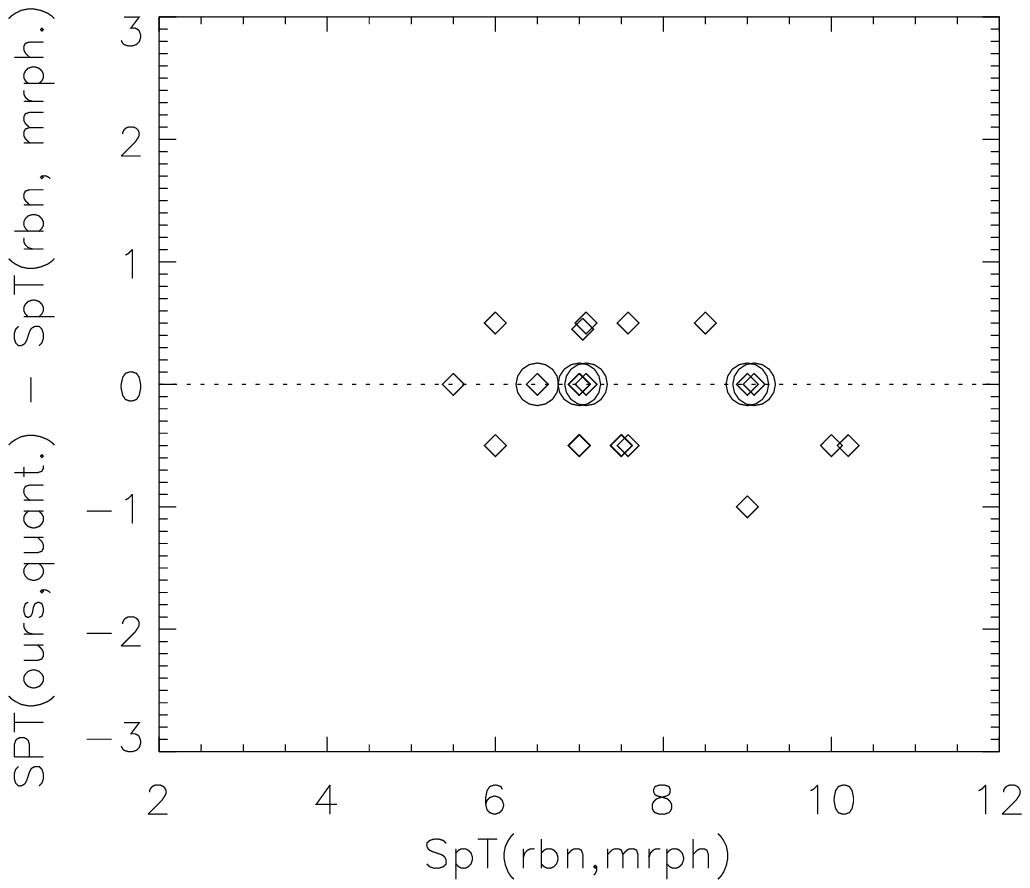}}
\resizebox{\hsize}{!}
{\includegraphics[width=6cm,height=4cm]{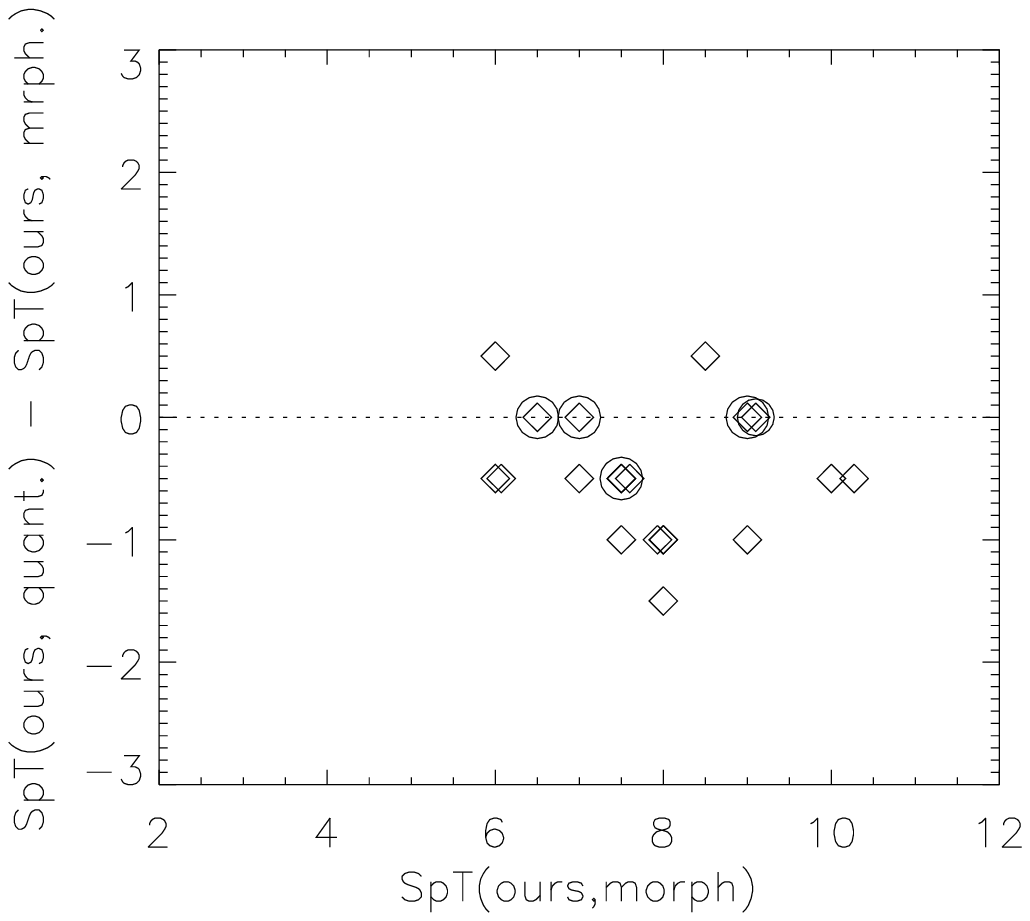}}
\caption{$Top$: High-resolution $quantitative$ versus  $morphological$ 
spectral types based on our high-resolution spectra degraded 
to $R$=4\,000.
$Bottom$ High-resolution $quantitative$ versus high-resolution 
$morphological$ spectral types.
}
\label{quant_old}
\end{figure}
It is important to verify the consisitency  of the two alternative 
schemes for the classification of O-type stars. 
Earlier studies \citep{CL74, mathys88, mathys89} have shown 
that $quantitative$ (either photographic or low-resolution 
digital) and $morphological$ classification based on photographic 
data are consistent to within {\it one to two subclasses}, with 
the former being systematically later (on average, by roughly 
half a spectral subtype).

The same comparison, but between the present $quantitative$ 
and the present $morphological$ classifications based on our degraded 
spectra ($R$=4\,000), indicates a closer agreement (within $\pm$0.5 
subtypes) without any systematic trend (Fig.~\ref{quant_old}, top panel). 
When considering the correspondence between the high resolution 
$quantitative$ and $morphological$ spectral types, we find 
evidence in our data of a systematic difference, where  
the former is now {\it earlier}, by up to 1.5 subtypes\footnote{This finding is 
a natural consequence of our results from Sect.~\ref{results}, 
namely that within the high-resolution spectra \heiia/ \heib\, can 
appear deeper than \heiib/\heiic, although their EWs are smaller.} 
(Figure~\ref{quant_old}, lower panel).

Thus, it appears that because of the higher quality of present-day 
optical spectra (which increases the accuracy of quantitative 
measurements and decreases the number of unresolved binary systems), 
the consistency between the two classification schemes can be 
significantly improved leading to practically equal results 
(provided the $morphological$ classification is performed 
using high-resolution spectra degraded down to $R$=4\,000).

\section{Summary and future perspectives.}
\label{conclusions}

Using high-resolution ESO/MTG FEROS spectra and applying the 
Walborn and the Conti schemes, we have performed a spectral 
classification of 19 Galactic O-type dwarfs and giants. We 
have investigated the correspondence between these high-resolution 
classifications and those attributed by Walborn \citep{walborn72, 
walborn73} based on low-quality photographic spectra and those 
assigned by \citet{mathys88} using low-resolution digital 
spectra. In addition, we have also investigated the spectroscopic 
status of our targets. The main results can be 
summarised as follows.

\paragraph{1. \it Spectroscopic binaries.}  
Using photometric/spectroscopic data from the literature in 
combination with results from a {\sc fastwind} model atmosphere 
analysis (Paper II), we confirm the binary nature of HD~152723 
and HD~152247, provide strong evidence of 
a companion in HD~93160 and HD~151003, and suspect possible 
duplicity in HD~69464. Four of these stars are members of 
cluster/associations and one is a field star. 
Thus, the binary fraction among our sample is 26 percent if 
both the cluster members and the field stars are considered. 
Given that the majority of our targets has never been monitored 
photometrically or spectroscopically, and that systems with 
orbital periods too long for an easy detection and too short 
for a direct angular separation can easily be lost, we 
consider this estimate as a lower limit.

For spectral classification, our analysis 
shows that binarity may or may not affect the derived 
spectral type/luminosity class, depending on the magnitude 
difference, separation, slit orientation etc. In any case,
and as might be expected, this effect does not lead to
systematic discrepancies (provided these systems are not used 
as standards), but `only' increases the uncertainty  
in the derived classification.

\paragraph{2. \it Correspondence between the present 
$morphological$ classification and the Walborn results.} 
Within a sample of 14 stars, we found evidence of 
a systematic discrepancy between our $morphological$ 
spectral types and those attributed by Walborn, with the 
former being later by up to 1.5 subtypes (Fig.~\ref{SPT_morph}). 
This discrepancy is larger than the corresponding 
uncertainties and thus significant. Concerning luminosity 
classes, no evidence of any systematic difference between 
our determinations and those from Walborn was found.
In our present understanding, the systematic discrepancy 
in $morphological$ classification is due to an interplay 
between stellar rotation and spectral resolution 
(Figures~\ref{res_rot} and \ref{spectra}) 
as well as technical differences in the classification 
process of photographic spectrograms and digital spectra
(Fig.~\ref{rbn}). 
Another important outcome of our analysis 
is that at least for narrow-lined stars (\vsini$\le$100~\kms) 
the ``rotational effect'' does not necessarily disappear when 
the high-resolution spectra are degraded to the
resolution of the Walborn standards 
(Figures~ \ref{res_rot} and \ref{resol_effect_2}). 

\paragraph{3. \it Correspondence between the present 
$quantitative$ classification and the Mathys results.} 
Within a limited sample of nine stars, 
we found that the spectral type criterion in the Conti 
scheme, log~$W'$, is sensitive to changes in the spectral 
resolving power (presumably due to effects on the EW of 
\heiic), while the luminosity class criterion, log~$W''$, 
is insensitive.  In particular,  the present high-resolution 
$quantitative$ spectral types tend to be systematically earlier 
(by typically half a subtype) than those attributed by Mathys. 
 Although comparable to the corresponding errors, the 
established discrepancy might be significant  because 
it is  systematic.

\paragraph{4. \it Correspondence between the present $quantitative$ 
and $morphological$ spectral types.}  Although different 
from earlier results, the modern-era $quantitative$ 
and $morphological$ classifications appear to be internally consistent 
(provided that the $morphological$ classification is performed 
using high-resolution spectra degraded to $R$=4\,000). 
This result, if confirmed by better statistics, might help 
us to 
improve the accuracy and consistency of  Galactic O star
classification after a corresponding new atlas of standard 
stars at $R$=4\,000 has been set up and published (see below).

\smallskip
\noindent

From the results outlined above, one might conclude 
that the $quantitative$ classification is more robust than the 
$morphological$ one. This conclusion, however, would be 
somewhat premature since it would be based on a very limited 
sample of stars that furthermore does not include supergiants.
On the other hand, a spectral type discrepancy of even half 
a subtype, if confirmed on the basis of higher quality statistics, 
might be significant and therefore require 
the Conti scheme to be updated using the power of high-quality 
spectral observations and line profile simulations. 
Such a thorough study is foreseen within our collaboration 
in the ''{\it VLT-FLAMES 
Tarantula Survey}" (PI: C.J. Evans). 

Concerning the $morphological$ classification, a 
possible solution of the problems identified in the 
present study requires an update of the corresponding 
approach. This revision might comprise the following 
steps: (i) selection and observation (at relatively 
high spectral resolution and S/N) of classification 
standards, which will be used to create a new digital 
atlas superseding the one of \citet{wf90}; 
(ii) development of a new classification technique that 
allows the spectra of both unknown and standard 
stars to be compared at the same resolution and  \vsini, and 
(iii) reclassification of as many Galactic O-stars as 
possible (observed uniformly and at the same $R$ as the 
standards) to be used as a firm basis for future studies.

First steps in this direction have been undertaken 
within the  ``{\it Galactic O-stars Spectroscopic Survey} 
(GOSSS)''  (PI: J. M. Appel\'aniz)  and the ''{\it VLT-FLAMES Tarantula 
Survey}''. 
Within the GOSSS survey, a digital atlas of Galactic O-type 
stars from both hemispheres will be created and published, 
to supersede the one of \citet{wf90} in terms of both the 
quality of the data (S/N $\approx$ 200, $R \approx$ 2\,500) 
and the number 
of standard stars included. Within the ``{\it VLT-FLAMES Tarantula 
Survey}'' project, on the other hand, we plan to publish a 
digital atlas of standard O-type stars using high-resolution ($R$
from about 40\,000 to about 80\,000), high signal-to-noise (S/N
$\approx$ 500) spectra collected from the ESO public archive and the
IACOB database \citep{simon10}. To diminish the combined effects of
resolution and rotation and ensure consistency with the modern-era 
$quantitative$ classification, the spectra will be degraded 
to 
$R$=4\,000, and subsequently used to create an atlas of unprecedented
quality.

Finally, we point out that although updated with respect 
to the resolution and rotational effects, the classification  
of O-type stars might still be subject to significant third parameter 
effects related to metallicity  (see \citealt{markova09} and references 
therein).

\acknowledgements{
We like to thank our anonymous referee for useful comments and 
suggestions. N.M. gratefully acknowledges travel grants by the 
IAC, Tenerife, Spain, and the Catania Observatory, Italy. This 
investigation was supported by the Bulgarian NSF (contract DO 02-85). 
SSD and AH acknowledge financial support from the Spanish Ministerio 
de Ciencia e Innovaci\'on under the projects AYA2008-06166-C03-01 
and the Consolider-Ingenio 2010. Program grant CSD2006-00070: 
First Science with the GTC  (http://www.iac.es/consolider-ingenio-gtc).
}

\appendix

\section{Comments on individual stars.}
\label{individual_stars}

\paragraph{HD~64568} $-$ This star was classified as O4~V((f))
\citep{GHS77}, O5~V \citep{loden, crampton72, P81}, and O3~V((f*))
\citep{walborn02}. The FUV/UV spectrum is also consistent with 
the O3~V((f*)) classification \citep{GB0404}. The visual appearance 
of our spectrum is identical to that shown in Walborn et al., 
suggesting the same O3~V((f*)) designation. 
The discrepant spectral types assigned at different observational 
epochs might indicate that HD~64568 is a binary.  This possibility 
seems to be supported by the detected \Vr - variability \citep{SN86}, 
but the good agreement between the model and the observed spectra 
(Paper II) argues in favour of the single-star hypothesis. The 
star has not been photometrically monitored and its short/long-term 
behaviour is completely unknown.  Lacking direct
evidence of possible binarity, we are inclined to consider HD~64568 as a
single object with \Vr - variations caused more likely by stellar 
pulsations (and/or wind variability). In this case then, the spectral
type discrepancy could be due to reasons other than spectroscopic
binarity.

\paragraph{HD~93204} is classified as O5~V((f)) by \citet{walborn73} 
and considered as a standard of this subtype. \citet{walborn02} repeated
this designation, whereas \citet{mathys88} proposed O5.5~V((f)). Our
high-resolution spectrum is identical to the one shown in 
\citet{walborn02}. Consequently, we should denote the star O5~V((f)), but
in our spectra, as well as in Walborn's, \heiia\, is of similar strength as 
\heiib, which is the definition of an O6 subtype. The measured 
log~$W'$~=~$-$0.38 and log~EW(4686)=$-$0.17 imply the O5.5~V classification.  
Within the limits of their $V$-band speckle interferometric observations, 
\citet{mason09} conclude that HD~93204 is a visually single object. 
To our knowledge, this star has never been monitored photometrically or 
spectroscopically. In Paper II, we have  no problems obtaining good 
fits to all strategic helium and hydrogen lines using one set of parameters 
(including \Vr and \vsini). Projected rotational velocities measured at 
different observational epochs agree perfectly (e.g., \vsini=130, 
\citealt{CE77}; \vsini=130~\kms, \citealt{penny96}).  All this suggests 
that HD~93204 is a spectroscopically single object. 

We note  that \citet{walborn10b} have recently reclassified HD~93204 
from O5((f)) to O5.5((fc)), where ``fc'' indicates the presence of \ciiie 
emission lines of comparable strength to those of \niiie.
Interestingly, in our high-resolution spectrum ($R$ about ten times higher 
than that in the Walborn et al. observations), no sign of any C~III 
emission can be seen, suggesting that the amount of C~III
emission is temporally variable in HD~93204. Similar results have been
reported for three stars of the Of?n category: HD~108, HD~191612, and
CPD~$-$28\,2561 \citep{walborn10b}. 
The physical reason for the variable C~III emission is still not well
understood but given that HD~108 and HD~191612 have been recognized as
long period binaries with 
magnetic fields, an interpretation in terms of orbital phase occultation of
a localized C~III emitting region has been suggested (\citealt{walborn10b}
and references therein). If confirmed by means of a more detailed
investigation, the single nature of HD~93204 would require an 
explanation of the variable C~III emission that is different from that suggested 
by \citet{walborn10b}.

\paragraph{HD~93843} $-$ This star was initially classified as O6 III(f)
\citep{walborn72}, and later on reclassified as O5 III(f) \citep{walborn73}.
Applying the quantitative approach, \citet{mathys88} assigned O5.5~III(f).
The morphology of our spectrum agrees well with the O6~III(f) designation
but with C~III in emission. Following the definition of the new Ofc
category (see above), we would denote it by O6~III(fc), that is one subtype
later than the designation by \citet{walborn10b}, which is O5~III(fc). 
The measured log~$W'$~=~$-$0.37 is consistent with the O5.5 subtype. 
In the interferometric survey of \citet{mason09}, HD~93843 is 
flagged as an object with  a ``null companion detection''. The star 
does not seem to show variations in its \Vr\, \citep{gies87}. The model 
atmosphere analysis does not reveal any discrepancy in the shifts 
and widths of strategic lines (Paper II). Individual 
\vsini~-~estimates derived in different observational epochs agree 
within the corresponding uncertainties (e.g., \vsini=120~\kms, 
\citealt{CE77}; \vsini=95~\kms, \citealt{howarth97}; \vsini=100~\kms, 
\citealt{penny96}). We suggest that HD~93\,843 is a spectroscopically 
single object.

\paragraph{CPD -59\,2600} $-$ \citet{walborn73} classified this star as
O6~V((f)). The morphology of our spectrum resembles that of HD~101190, the
primary standard for the O6~V((f)) subtype \citep{wf90}, and consequently we
assign O6~V((f)) to our target as well; log $W'$~=~$-$0.17 and 
log~EW(4686)~=~$-$0.24 indicate O6.5~V/III. 
A literature research shows that CPD -59\,2600 has never been observed
systematically.  A comparison of individual \vsini-estimates does not show
any significant variations in this parameter (e.g., \vsini= 140~\kms,
\citealt{CE77}; \vsini=142~\kms, \citealt{howarth97}; \vsini=135~\kms, 
\citealt{penny09}). This finding together with the good agreement between
synthetic and observed strategic lines (Paper II) suggests that CPD~-59\,2600
is likely a single object.

\paragraph{HD~63005} was classified as O7 by \citet{loden} and 
O6~V((f)) by \citet{GHS77}. The general appearance of our spectrum agrees
well with the O7.5 designation: \heib\, is slightly deeper than 
\heiic, and there are both weak
N~III lines in absorption and  missing C~III~$\lambda$4650 absorption. 
Since  \heiid\, is comparable in strength to \heiic, the star might be
 a dwarf, but the strength
of the \niiie emission lines are more consistent with a luminosity
class ``III'' rather than ``V'' designation. Based on these 
considerations, we adopt O7.5~V((f)), but draw attention to its 
somewhat strong N~III emission. The measured log~$W'$~=~$-$0.13 
and log~EW(4686)~=~$-$0.12 indicate O6.5~V.
No reports about duplicity were found in the literature. 
The model atmosphere analysis has not revealed any spectral 
discrepancy (Paper II). Projected rotational velocities estimated 
at different epochs agree perfectly (e.g., \vsini=73~\kms, \citealt{penny96}; 
\vsini=74~\kms, \citealt{howarth97}; \vsini=80~\kms, \citealt{penny04}). 
With all this in mind, we consider HD~63005 a spectroscopically single 
object.

\paragraph{HD~152723, SB1} $-$ This star was classified as O7 by
\citet{schild} and O6.5~III(f) by \citet{walborn72} and  \citet{GHS77}.
The appearance of our spectrum is consistent with the O6.5~III(f)
designation: \heiia\, is slightly deeper than \heiib; \heib\, and 
\heiid are weaker than \heiic, and there is weak N~III emission. 
The classification determined by means of the measured 
log$ W'$=$-$0.12 and log~EW(4686)~=~$-$0.36 is also 
O6.5~III.
\citet{mason09} resolved HD~152723 as a visual binary with 
an angular separation of 0.098''. Based on observed \Vr~- variability, 
\citet{full96} suggested that HD~152\,723 is a SB1. This view was 
supported by \citet{lefevre09} who discovered photometric variations 
with a period of 0.395$^{d}$. The discrepant strengths and widths of 
strategic helium and hydrogen lines (Paper II) also argue in favour 
of the binary hypothesis. Individual \vsini - estimates available in 
the literature however do not deviate  significantly, e.g., 
\vsini=110~\kms, \citealt{balona75}; 
\vsini=130~\kms, \citealt{CE77}; \vsini=111~\kms, \citealt{howarth97}; 
\vsini=123~\kms, 
\citealt{penny96}. We consider HD~152723 as a single-lined SB.

\paragraph{HD~93160, SB1} $-$ \citet{walborn72} denoted this star by
O6~III(f). From the visual inspection of our spectrum, we classify it as
O7~III(f): \heib\, of similar strength as \heiic; weak \heiid absorption in 
combination with weak \niiie emission. The quantitative diagnostics, log$
W'$~=~$-$0.05 and log~$W''$~=~0.05, indicate a less luminous object, namely a
dwarf, of the same O7~subtype.
The spectral type discrepancy between the present $morphological$ and
Walborn's classification might indicate that HD~93160 is a spectroscopic
binary. This possibility was initially suggested by \citet{gies87} based on
\Vr~-~arguments, the latter however being questioned by \citet{levato91}.
Our model atmosphere analysis has revealed a number of discrepancies in the
optical spectrum, which might be interpreted as an indication of a 
close companion (Paper II). The binary hypothesis is supported by 
the large differences in \vsini\, derived in different observational
epochs: e.g., \vsini=180~\kms, \citep{CE77}; \vsini=205~\kms,
\citep{UF}; \vsini=145~\kms (Paper II). This star is very likely a SB1.  

\paragraph{HD~94963} was previously classified as O6.5~III(f) by 
\citet{walborn73} and  O7~III by \citet{GHS77}. Our spectrum of 
this star agrees with the O7.5~II subtype. Within the corresponding 
error, log~$W'$~=$-$0.01 is consistent with the O7-7.5 subtype, while 
log~$W''$~=~0.32 corresponds to a supergiant or at least a giant. 
The discrepant classification attributed by various authors at 
different epochs might indicate that HD~94963 is a spectroscopic 
binary. This possibility seems to be supported by 
the large spread in individual \vsini - estimates available in the 
literature (e.g., \vsini=130~\kms, \citealt{UF}; \vsini=90~\kms, 
\citealt{CE77}). However, the star is regarded as photometrically 
stable \citep{kilkenny98}, with a constant \Vr\, \citep{gies87}. 
In addition, 
the model atmosphere analysis does not reveal any discrepancy except 
for one: the \heiid\, absorption line is peculiarly weak and narrow 
with a core that is red-shifted with respect to the position of the 
other spectral lines (Paper II). If not caused by wind asymmetries, 
the 
latter result might indicate a composite profile, consisting of an 
emission feature superimposed on the blue wing of an absorption profile. 
At least at present, we consider HD~94963 a single object 
with an asymmetric wind.

\paragraph{CPD -58\,2620} was classified as O6.5~V((f)) by 
\citet{walborn73} and O8~III by \citet{morrell88}. The morphology of our spectrum is
similar to that of $\lambda$~Ori, the classification standard for
O8~III((f)) \citep{wf90}. However, in $\lambda$~ Ori as well as in 
CPD~-58\,2620 the \niiie emission lines are very weak, and \heiid\, is deeper
than \heiic, suggesting the ``V((f))'' rather than ``III(f)'' designation.
Thus, we classify this star as O8~V((f)). From the measured 
equivalent widths, we estimate log~$ W'$=$-$0.07 and log~$W''$~=~0.26,
which is indicative of O7~III. 
CPD~-58\,2620 is a member of a visual binary with 3.7 arcsec separation. The
star has been recognised ``as a possible radial-velocity variable'', 
which
however ``does not show any obvious orbital trend in velocity from night to
night'' \citep{penny93}. No information about systematic photometric 
variations was found in the literature but a comparison of several $UBV$
estimates listed in the Reed (2005) catalogue (V/125) suggests that 
this star
might be photometrically variable. The excellent agreement between
theoretical and observed strategic lines suggests that it is more 
likely to be a
single object. Lacking any evidence of binarity, we consider 
CPD~-58\,2620 as a spectroscopically single object, but draw attention to the
discrepancies in luminosity class between the various spectral
diagnostics.

\paragraph{HD~69464, SB2?} $-$ \citet{walborn72} classified this 
star as O6.5~Ib(f), and defined it as a classification standard 
for this spectral type/luminosity class \citep{wf90}. \citet{GHS77} 
attributed O7~III(f). The visual appearance of our spectrum is 
similar to that of the spectrum shown in \citet{wf90}, with one 
important difference: in the latter, \heiid\, is completely 
filled in by wind emission, while in our spectrum it appears as a 
weak absorption feature. Based on this notion and accounting for 
\heib\, being somewhat deeper than \heiic\, in our 
spectrum as well as in that of \citet{wf90}, we re-classify 
the star from O6.5~Ib(f) to O7.5~II(f). Within the corresponding 
error, log~$ W'$~=$-$0.01 indicates the O7-7.5 subtype.
In contrast, log~$ W''$~=~0.28 is consisitent with the star 
being a giant. 
Spectral anomalies such as those outlined above are usually 
interpreted as an indication of a companion. Unfortunately, 
HD~69464 has never been observed systematically, 
and no direct evidence in support or against this possibility 
exists. Individual \vsini\, determinations (e.g., \vsini=71~\kms, 
\citealt{howarth97}; \vsini=82~\kms, \citealt{penny96}) and 
\Vr - estimates (e.g., \Vr=43~\kms, \citealt{crampton72}; \Vr=45~\kms, 
Paper~II ) seem to agree well, but this may still be due to 
observational selection. 
The correspondence between synthetic and observed lines is also good, 
except for \heiid: the observed feature appears peculiarly 
narrow and weak and may show an inverse P Cygni profile (Paper II). 
If not due to wind variability or asymmetries, the latter finding 
would suggest that HD~69464 is a SB2. Systematic spectral/photometric 
observations are needed to check this possibility in the future.  

\paragraph{HD~93222} $-$ Classified as O7~III((f)) by
\citet{walborn71, walborn72}, and again by \citet{levato81} and 
\citet{wf90}. However, \citet{massey01} attributed O8~III((f)). 
The appearance of our spectrum differs significantly  from 
that shown in \citet{wf90}, and resembles much more the
spectral morphology of the O8~III((f)) standard $\lambda$~Ori, 
which we adopt in our classification. The measured log~$W'$~=$-$0.03 
and log~$ W''$~=~0.28 indicate a slightly earlier subtype, namely O7~III. 
HD~93222 is considered as non-variable in both photometry (e.g.,
\citealt{moffat77, moffat78}) and \Vr (e.g., \citealt{levato90}). \vsini -
estimates from different observational epochs agree quite well (e.g.,
\vsini=65, \citealt{CE77}; \vsini=77~\kms, \citealt{howarth97};
\vsini=77~\kms, \citealt{penny96}). The model atmosphere analysis does not
uncover any spectral discrepancy (Paper II). All this suggests that
HD~93222 is a spectroscopically single star.

\paragraph{HD~91824} $-$ \citet{walborn72} denoted this star 
by O7~V((f)), confirmed by \citet{GHS77}. From a visual inspection 
of our spectrum, we suggest O8~V((f)) since \heiid\, and \heib\, 
are deeper than \heiic; numerous N~III lines are in absorption,
and weak C~III in absorption and weak \niiie are in emission. 
The measured 
log~$ W'$ =$-$0.11 and log~EW(4686)~=~$-$0.05 indicate O6.5~V. 
HD~91824 was found to be an irregular variable in both photometry
\citep{lefevre09} and  \Vr\, \citep{feast58}. Since the model 
atmosphere analysis (Paper II) as well as the classification analysis 
(present study) did not reveal any spectral discrepancy, we suggest 
that this variability is more likely due to stellar pulsation than 
to binarity.  This possibility is further supported by the good 
agreement between individual \vsini - estimates: \vsini~=~67~\kms 
\citep{penny96}  and \vsini=65~\kms \citep{howarth97}. 

\paragraph{CD~-43\,4690} $-$ Classified as O7.5 III(f) by
\citet{crampton71} and thereafter by \citet{mathys88}. The morphology 
of our spectrum agrees well with the O7~III(f) designation: 
\heib\, is as deep as \heiic; N~III is in emission, and  \heiid\, 
is somewhat weaker than \heiic. We measure log~$ W'$~=$-$0.20 and 
logEW(4686)~=~$-$0.23,  
which is indicative of the O6.5~V/III type. 
This star has never been observed systematically, neither 
photometrically nor spectroscopically. A comparison of individual 
data provided in the Reed (2005) catalogue did not reveal any 
significant variations in $V$ and $B-V$ on a longer (year) 
timescale.  Based on several snapshots observations, \citet{crampton72} 
conclude that this star is not a \Vr - variable.  The model atmosphere 
analysis does not reveal any spectral discrepancy (Paper II). We 
consider CD~-43\,4690 as a spectroscopically single object.

\paragraph{HD~92504} was denoted by O8.5~V \citep{walborn73}, 
and this classification was repeated by \citet{GHS77}. 
\citet{turner77} assigned O9~V. The visual appearance of our 
spectrum is identical to that of the O8.5~V standard HD~46\,149 
\citep {wf90}. We measure $\log W'$~=~0.32 and $\log W''$~=~0.11, 
indicative of the O9~III/V type. 
Individual $UBV$ measurements listed in the Reed (2005) catalogue 
suggest that the star is photometrically stable on a long (year) 
timescale. Individual \Vr - estimates given in various catalogues 
(e.g., III/190 - \citealt{DFM95}; V/125 - \citealt{reed05}) agree 
well. The model atmosphere analysis has not revealed any discrepancy 
in the spectrum (Paper II). This star is likely a single object.

\paragraph{HD~151003, SB1?} was classified as O9~II \citep{walborn73},
O9.5~III \citep{mathys88} and O9~Ib \citep{GHS77}. The morphology of our
spectrum is in-between that of the classification standards HD~148\,546
(O9~Ia) and $\iota$~Ori (O9~III) \citep{wf90}, suggesting an intermediate 
O9~II classification. log~$W'$~=~0.32 and log~$W''$~=~0.24 indicate O9~III. 
The differences in the spectral classifications of HD~151003 
might indicate that it has a composite spectrum caused by binarity. 
Within the limits 
of the Mason et al. $V$-band interferometry, no indication of any 
companion was found. In the \citet{turner08} adaptive optics $I$-band 
photometric survey the star is flagged with ``V'', i.e.,
''one or more additional spectroscopic companions identified''. Based on \Vr
- arguments, \citet{gies87} suspected that HD~151\,003 is a SB1. This view is
supported by discrepant widths of strategic helium and Balmer lines
(Paper II).  Thus, it seems likely that HD~151\,003 is an SB1.  Systematic
photometric and/or radial velocity observations are needed to convincingly
prove this possibility.  

\paragraph{HD~152247, SB2} $-$ According to \citet{walborn73}, this star
is a O9.5II/III and, similarly, a O9.5~III following \citet{mathys88}. Our
spectrum is almost identical to that of HD~151003, suggesting the same
O9~II classification. The measured log~$ W'$~=~0.39 is consistent with the
O9 subtype; log~$W''$~=~0.19 corresponds to the luminosity class ``III''
rather than ``II''. 
\citet{balona83} reported that the visual brightness of HD~152247 
is stable to within $\pm$0.05 mag. The star has never been monitored 
spectroscopically, but a comparison of individual estimates from 
the literature shows that its \vsini\,  appears to be stable (e.g., 
\vsini=110~\kms \citealt{CE77}; \vsini=120~\kms \citealt{penny96};
\vsini=112~\kms \citealt{howarth97}), while its \Vr\,  
varies \citep{Rab96, sana08}. On the basis of the detection of an 
additional weak absorption component in \heiid, \citet{sana08} 
suggested that HD~152247 is a SB2 consisting of an O9~III and an 
O9.7~V component. This result is additionally supported by detected 
discrepancies in the
widths of helium and Balmer lines (Paper II). HD~152427 is definitely a
SB2, where the secondary is significantly fainter (by 3 or more magnitudes
to be consistent with the ``null companion detection'' of \citealt{mason09})
than the primary.

\paragraph{HD~302505} $-$ Classified as O9.5~III by \citet{gg70}
and O8.5~III by \citet{GHS77}. \citet{crampton72} listed it as B2
without, however,  any reference to the original source.  
Our spectrum,  visually, resembles that of $\iota$~Ori, the 
classification standard of the O9~III type \citep{wf90}. 
log~$ W'$~=~0.16 and log~$ W''$~=~0.22 indicate O8~III.  
The lack of consistency between the assigned spectral types might 
be interpreted as evidence  of a companion.  However, 
this view is not supported by the close agreement between the model 
and the observed spectra (Paper II). Individual $UBV$ measurements 
listed in the Reed (2005) catalogue suggest that HD~302505 is 
photometrically stable on a longer timescale (within $\Delta V \le$~0.1~mag, 
$\Delta(B-V)~\le$~0.1~mag). Thus, we consider this star
a single object, but future photometric and radial-velocity monitoring is
required to check this view.

\paragraph{CPD~-44\,4865} has been classified as O9.5~Ib by \citet{feast61};
\citet{mathys88} reclassified it as O9.7~III. The general appearance of our
spectrum resembles that of HD~48434, the classification standard for
B0~III \citep{wf90}. The only noticeable difference is that in
CPD~$-$44\,4865, the lines \heiib\, and \heiic\, are definitely 
present, while they are missing in HD~48434. 
log~$W'$~=~0.61 and log~$W''$~=0.19 indicate O9.5~III, 
in good agreement with the classification by \citet{mathys88}.
To our knowledge, CPD~-44\,4865 has never been systematically observed. 
In the Reed catalogue (2005), it is flagged as "RV var", but no 
reference to the corresponding source(s) is provided. A comparison of 
individual \Vr - estimates from different observational epochs does 
not show any significant variations (e.g., \Vr~=~40~\kms, 
the GCMRV; \Vr=39.3~\kms, the Reed 2005 catalogue). The model atmosphere 
analysis has not posed any problem: excellent fits were obtained for 
all strategic lines with one set of parameters (Paper II). All this 
suggests that CPD~-44\,4865 is likely a spectroscopically single object.

\begin{table*}
\begin{center}
\caption[c]{Equivalent widths (in \AA) and equivalent width ratios 
as measured for the O-type stars in our sample. Accuracy of 
individual equivalent width measurements: 0.02 to 0.07 dex, depending 
on the strength of the line (the stronger the line, the higher the 
accuracy). $Quantitative$ classification accounting for the error 
in log~$W'$ ($\sim\pm$ 0.03 dex) and log~$W''$($\le\pm$0.05 dex).
} 
\label{EW}
\tabcolsep1.0mm
\begin{tabular}{lcccccrcl}
\hline
\hline
\multicolumn{1}{l}{Star}
&\multicolumn{1}{l}{\heib}
&\multicolumn{1}{l}{\heiic}
&\multicolumn{1}{c}{\heiaa}
&\multicolumn{1}{l}{Si~IV~$\lambda$4089}
&\multicolumn{1}{c}{log EW(4686) $^{a}$}
&\multicolumn{1}{l}{log W' $^{b}$}
&\multicolumn{1}{l}{log W'' $^{c}$}
&\multicolumn{1}{l}{quant. class.}
\\
\hline                                
HD~64568            &--     &0.84   & --  &  -- &-0.46 &        &     & \\
HD~93204            &0.34   &0.81   & --  &  -- &-0.17 &$-$0.38 &     &O5.5 V \\
HD~93843            &0.30   &0.70   & --  &  -- &-0.72 &$-$0.37 &     &O5.5 III \\
CPD~$-$59\,2600     &0.43   &0.63   & --  &  -- &-0.24 &$-$0.17 &     &O6.5 V/III\\
HD~63005            &0.59   &0.80   & --  &  -- &-0.12 &$-$0.13 &     &O6.5 V \\
HD~152723           &0.45   &0.60   & --  &  -- &-0.36 &$-$0.12 &     &O6.5 III\\
HD~93160            &0.55   &0.62   &0.15 &0.17 &      &$-$0.05 &0.05 &O7 V\\
HD~94963            &0.59   &0.61   &0.11 &0.23 &      &$-$0.01 &0.32 &O7-7.5 I/III\\
CPD~$-$58\,2620     &0.64   &0.76   &0.12 &0.22 &      &$-$0.07 &0.26 &O7 III\\
HD~69464            &0.64   &0.66   &0.10 &0.19 &      &$-$0.01 &0.28 &O7-7.5 III\\
HD~93222            &0.60   &0.65   &0.12 &0.23 &      &$-$0.03 &0.28 &O7 III\\
HD~91824            &0.58   &0.75   &0.13 &0.15 &-0.05 &$-$0.11 &     &O6.5 V  \\
CD\,$-$43\,4690     &0.47   &0.75   &0.08 &0.16 &-0.23 &$-$0.20 &     &O6.5 V/III\\
HD~92504            &0.87   &0.42   &0.29 &0.37 &      &0.32    &0.11 &O9 III/V \\
HD~151003           &0.78   &0.37   &0.25 &0.44 &      &0.32    &0.24 &O9 III   \\   
HD~152247           &0.86   &0.35   &0.30 &0.47 &      &0.39    &0.19 &O9 III \\
HD~302505           &0.79   &0.55   &0.22 &0.37 &      &0.16    &0.22 &O8 III \\
CPD\,$-$44\,4865    &0.89   &0.22   &0.32 &0.50 &      &0.61    &0.19 &O9.5 III \\
HD~69106            &0.95   &0.17   &     &     &      &0.74    &     &O9.7     \\
\hline
\end{tabular}
\end{center}
$^{a} -$ log~EW($\lambda$ 4686) determines the luminosity class 
for stars of spectral types O6.5 and earlier (see \citealt{mathys88}).\\
$^{b}$ $-$ log~$W'$~=~$\log EW$ (HeI$\lambda$4471) $-$ $\log EW$(HeII$\lambda$4541) 
determines the spectral type (see \citealt{conti71} and \citealt{mathys88}).\\
$^{c} -$ log~$W''$~=~$\log EW$ (HeI$\lambda$4143) $-$ $\log EW$ (SiIV$\lambda$4089) 
determines the luminosity class for stars of O7 and later \citep{conti71}.
\end{table*}

\paragraph{HD~69106} $-$ \citet{morris61} classified this star as
B0.5~III; \citet{GHS77} reclassified it as B0.5 IVnn. Our spectrum
resembles that of $\tau$~Sco, the classification standard for B0.2~V
\citep{wf90}. The measured log~$W'$=0.74 corresponds to O9.7.
A literature research shows that HD~69106 has been recognised as a
photometrically variable star, with a period of 1.48~d
\citep{marchenko98}. The star has not been monitored spectroscopically.
Individual \vsini - estimates agree well (e.g., \vsini=329~\kms,
\citealt{howarth97}; \vsini=316~\kms, \citealt{balona75}; \vsini=328~\kms,
\citealt{CE77}). Excellent fits have been derived for all strategic lines in
the spectrum (Paper II). In the absence of direct evidence of binarity, we
suggest that HD~69106 is more likely a single very late O-type dwarf
undergoing stellar pulsations.

\section{High-resolution atlas of our sample of stars}
\begin{figure*}
\begin{center}
{\includegraphics[width=18cm,height=21cm]{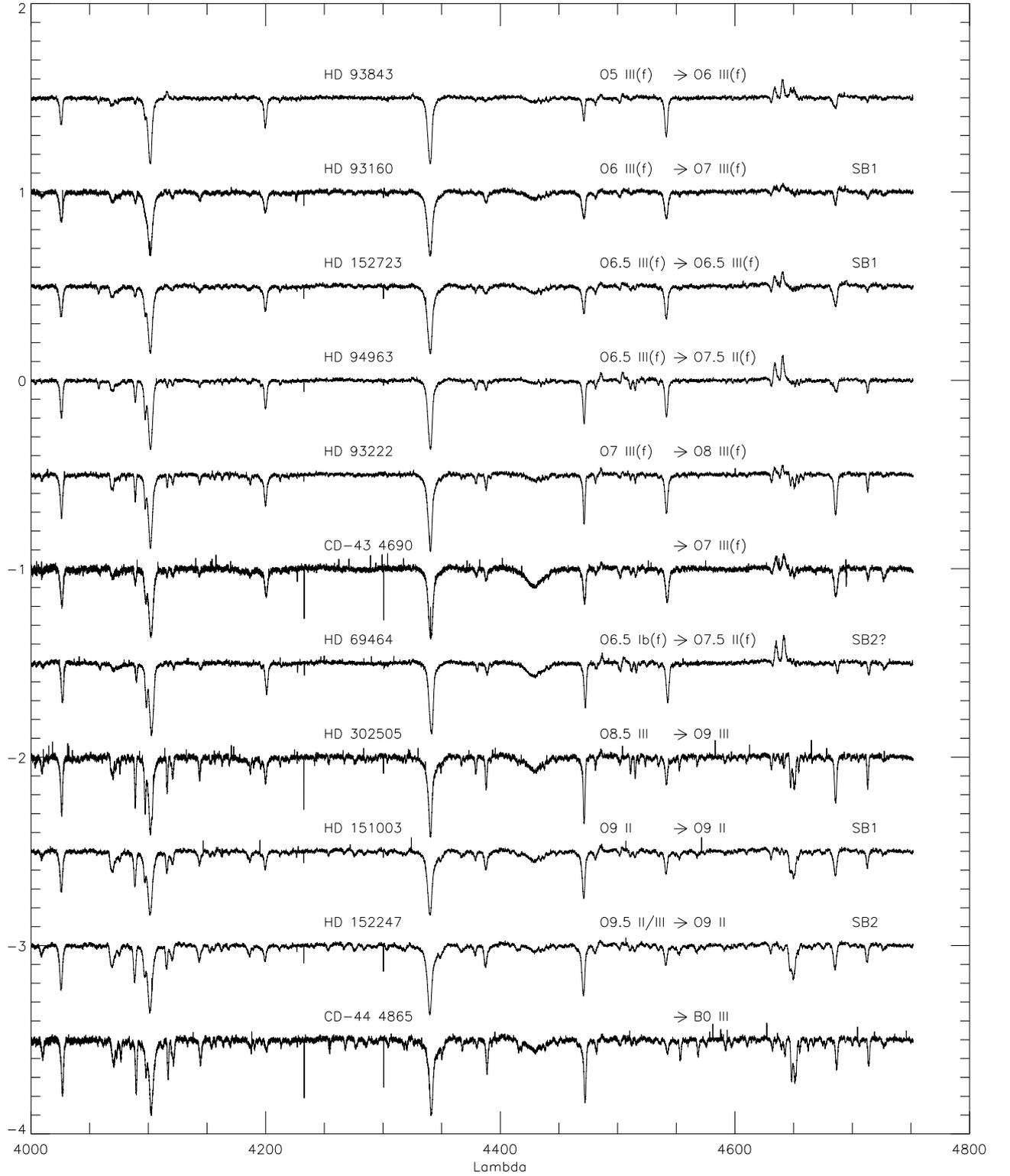}}
\caption{High-resolution spectra of sample stars which have been classified
originally as giants. $Morphological$ classification as provided by Walborn
\citep{walborn72, walborn73} using low resolution photographic spectra 
(with individual data from \citet{GHS77}), 
together with our $morphological$ classification based 
on high-resolution spectra. Confirmed/suspected SB1 and SB2 are 
also indicated.}
\label{gs_atlas}
\end{center}
\end{figure*}
\begin{figure*}
\begin{center}
{\includegraphics[width=18cm, height=15cm]{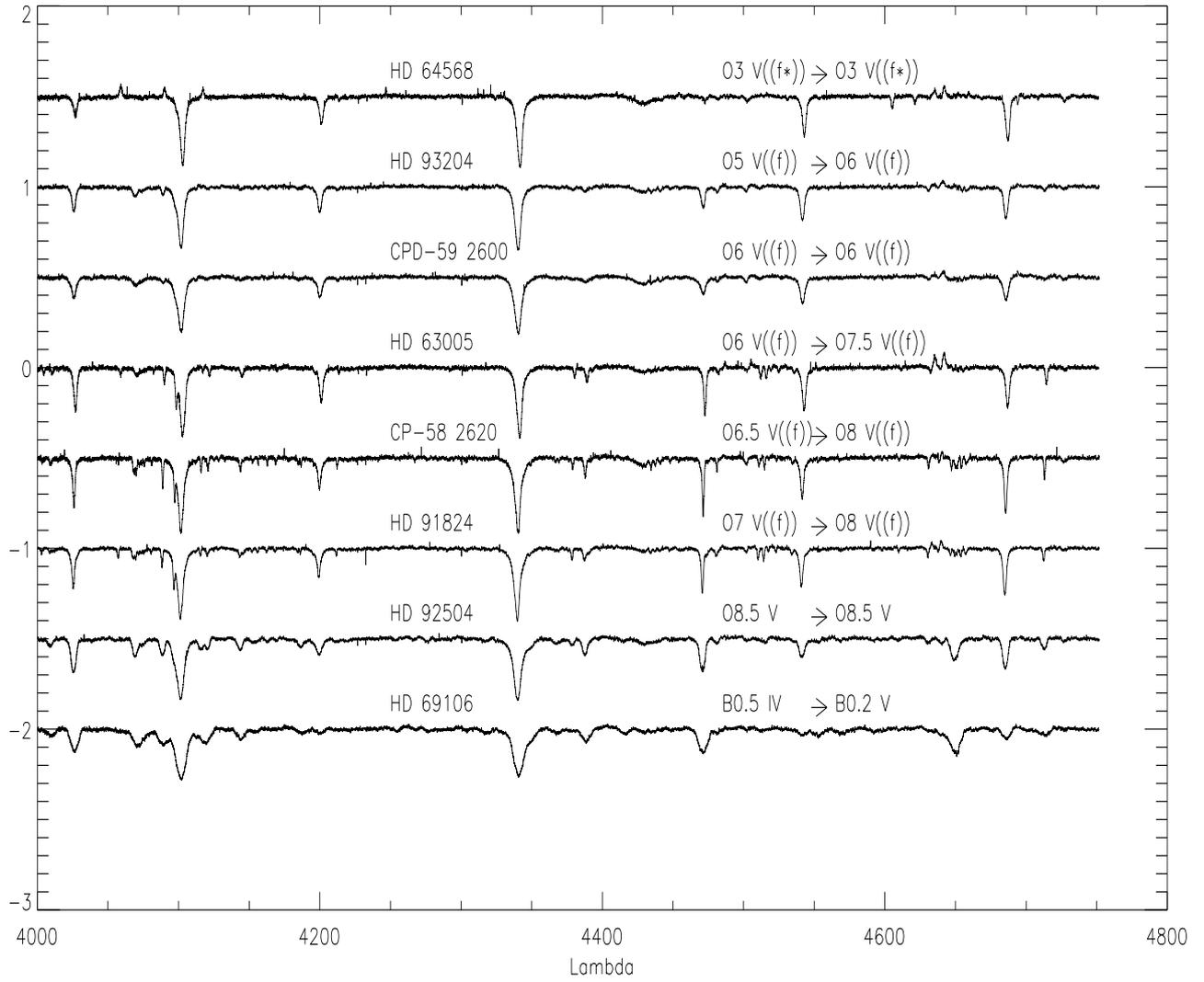}}
\caption{As Fig.~\ref{gs_atlas}, but for stars originally classified as
dwarfs.} \label{dw_atlas}
\end{center}
\end{figure*}
\end{document}